\documentclass{article}

\usepackage[a4paper, top=2cm, bottom=2.5cm, left=3cm, right=3cm]{geometry}
\usepackage{graphicx} 
\usepackage{url}  
\usepackage{hyperref}
\usepackage{subcaption}
\usepackage{authblk}
\usepackage{amsmath}
\usepackage{amssymb}
\usepackage{multirow}
\usepackage{natbib}
\providecommand{\keywords}[1]{\textbf{Keywords: } #1}

\title{Evaluating the Prediction of Wind Power Ramping Events in the Belgian Offshore Zone}

\author[1,2]{Ruoke Meng}
\author[1]{Geert Smet}
\author[1]{Dieter Van den Bleeken}
\author[3]{Aaron Van Poecke}
\author[1,3,4]{Hossein Tabari}
\author[3]{Peter Hellinckx}
\author[1,2]{Piet Termonia}
\author[1,3]{Joris Van den Bergh}

\affil[1]{Royal Meteorological Institute of Belgium, Brussels, Belgium}
\affil[2]{Ghent University, Ghent, Belgium}
\affil[3]{M4S, Faculty of Applied Engineering, University of Antwerp, Belgium}
\affil[4]{United Nations University Institute for Water, Environment and Health, Hamilton, ON, Canada}

\date{}

\begin{document}

\maketitle

\begin{abstract}
This study provides a comprehensive evaluation for the prediction of wind power ramping events in the Belgian Offshore Zone. These rapid, large-scale power fluctuations pose significant challenges to grid reliability. The research uses operational Numerical Weather Prediction (NWP) models from Royal Meteorological Institute of Belgium, as well as its version enhanced with Wind Farm Parameterization (WFP). Power predictions are generated with both typical power curves and machine learning approaches. Standard verification metrics, such as Mean Absolute Error (MAE), often fail to capture the operational significance of ramp events. To address this, we develop a flexible verification framework designed to assess ramp forecast performance. This framework incorporates adjustable time and power buffers, which tolerate minor, operationally acceptable discrepancies in the timing and magnitude of predicted events. Application of this framework to both intraday and day-ahead forecasts reveals that WFP-enhanced models consistently improve ramp predictions over the operational baseline. Further analysis reveals that while the WFP model with power curves effectively reduced false alarms, it comes at the cost of more misses. In contrast, ML-based approaches achieve slightly higher overall skill scores by striking a better balance between reducing these error types. Moreover, we introduce the Ramp Alignment Score (RAS), an event-based metric that quantifies the temporal alignment between predicted and observed ramps, to supplement the model evaluation by lead time. RAS analysis demonstrates that WFP models achieve better temporal alignment and reveals a distinct diurnal cycle in ramping prediction errors. Finally, we investigate the impact of a specific meteorological driver, finding an association between severe precipitation and large, highly predictable ramp events. Conversely, moderate and light precipitation are linked to a higher incidence of missed events and false alarms. This work provides both an operationally relevant evaluation methodology and insights into ramp predictions under specific meteorological conditions.
\end{abstract} 


\section{Introduction}
The integration of offshore wind power into energy grids is vital for the transition to sustainable energy systems. As a critical component of the Belgian national energy system, the Belgian Offshore Zone (BOZ) requires effective decision-making to ensure reliable energy scheduling and system operation. Wind power ramping events, characterized by rapid and significant changes in power generation, pose substantial challenges to operational energy management. Concerning the high-density wind farms installed at the BOZ \cite{murcia2020power}, the Belgian Transmission System Operator (TSO), Elia, reports potential interests in short-term ramping as these rapid power fluctuations challenge their operations \cite{elia2018offshore, elia2026adequacy}. During storms, up and down ramps involving nearly 50\% of the installed capacity can occur within 30 minutes, threatening grid stability and complicating real-time power balancing. Accurate prediction of such ramping events is a critical focus for both operational forecasting systems and renewable energy research.

Numerical Weather Prediction (NWP) models deliver key meteorological inputs for wind power forecasting, particularly for lead times longer than about 6 hours in the future \cite{kariniotakis2017renewable}. However, forecast accuracy remains limited due to the complex interactions between atmospheric processes and site-specific factors such as turbine wake effects and wind farm layout \cite{gonzalez2012wake, archer2018review}. To address these challenges, recent efforts have focused on improving NWP models through the integration of Wind Farm Parameterization (WFP), which represents the aerodynamic impacts of wind farms, such as wake effects and altered boundary layer dynamics, within the model's physics schemes \cite{fischereit2022review, van2022one}. A popular WFP solution is Fitch's method, which presents the effects of wind turbines by imposing a momentum sink on the mean flow and transferring kinetic energy into electricity and turbulent kinetic energy \cite{fitch2012local}. Recently, the Royal Meteorological Institute of Belgium (RMI) has incorporated this WFP approach into the operational Limited Area Model (LAM) ALARO-4km NWP model, and demonstrated improved wind forecast performance when validated against lidar measurements \cite{dieter2025improving}. 

While accurate wind speed forecasts are crucial for power prediction, ramping events remain difficult to forecast due to the inherently multi-scale and dynamic nature of atmospheric drivers \cite{greaves2009temporal}. These events may be triggered by diverse mechanisms, including frontal systems, turbulence, non-frontal precipitation, or turbine-related effects such as cut-off and mechanical shutdowns \cite{drew2018identifying, pichault2020characterisation}. Moreover, the causes of ramping events are often site- and case-specific, making it difficult to establish generalizable patterns across different wind farm settings \cite{gallego2015review}. Although WFP has demonstrated improvements in wind speed forecasts and benefits in power predictions, its performance particularly in the predictability for ramping events is still unclear. 

A comprehensive understanding of ramping predictability requires verification methodologies to evaluate model skill. Two key aspects of ramping forecast error must be addressed: timing errors, where the intensity of the ramp is forecast correctly but at the wrong time; and magnitude errors, where the timing is correct but the predicted magnitude differs from the observed value \cite{potter2009potential}. These error types highlight the challenge of simultaneously capturing both the temporal and amplitude characteristics of power ramps. Although Root Mean Square Error (RMSE) is commonly used to quantify forecast errors at each lead time and is also adopted in some ramping verification frameworks \cite{cutler2007detecting, jin2024evaluation}, this metric has been criticized for being insufficiently sensitive to the timing and structural aspects of ramps \cite{vallance2017towards, messner2020evaluation}.

The definition and verification of ramping events remain highly subjective and context-dependent \cite{bianco2016wind}. Many studies adopt threshold-based definitions and identify whether a ramping event exists based on preset magnitude and duration thresholds \cite{Kamath5484508,bradford2010forecasting}, and use contingency tables to evaluate forecasts based on binary outcomes, including hits, misses, and false alarms \cite{zhang2017ramp, cui2023algorithm}. This binary approach has been criticized for its drawback: high sensitivity to arbitrary but restrictive thresholds (e.g., a significant power change of 45\% may be ignored if the threshold is set to 50\%) \cite{gallego2013wavelet}. Alternative approaches, including wavelet transforms \cite{cheneka2020simple} and edge-based metrics \cite{bossavy2015edge}, have been explored to discuss the temporal consistency of ramp detection. Metrics such as Dynamic Time Warping (DTW), which measures the overall shape similarity between two time series through non-linear alignment \cite{zhang2022short}, and the Trend Direction Index (TDI), which assesses the directional consistency (i.e., upward or downward trends) between forecasted and observed sequences \cite{frias2016introducing}, have been employed to evaluate ramping predictions. These metrics, although mathematically rigorous, may not provide an intuitive understanding of ramp event predictability, especially in operational decision-making contexts where binary outcomes are more accessible to stakeholders.

To provide an understanding of the prediction skill of ramping events, we perform ramping forecast verification based on the WFP enhanced NWP model, in comparison with the RMI's operational NWP models and power conversion methods, including power-curve conversion and machine learning. We present a comprehensive and easy-to-use framework to enhance ramping verification methods and to achieve a fair comparison among different forecasting models. Our approach aims to address the variability in wind power predictions and improve the flexibility of ramping event verification. 

This paper discusses topics of power prediction and corresponding ramping verification with the following structure: (i) Several wind power models are evaluated by their bias and Mean Absolute Error (MAE) to offer a general view of power prediction accuracy. (ii) The corresponding ramping forecasts are investigated by their MAE and variability, and the limitations of MAE analysis for ramping verification are discussed. In response to these shortcomings,(iii) a flexible ramping event verification framework is introduced, which incorporates time and power buffers. It is then applied to evaluate models' skill under multiple buffer configurations for intraday and day-ahead forecasts. Building on this, (iv) a Ramp Alignment Score (RAS) is proposed to quantify the temporal alignment between predicted and observed events within defined time windows. Finally, to demonstrate the explanatory function of these approaches, (v) we investigate the prediction skill by categorizing ramping events according to precipitation intensity. This work introduces approaches tailored to ramping analysis, which overcome the limitations of standard metrics for ramping verification, and contribute to the understanding of ramping predictions under specific meteorological conditions.

\section{Data}
\subsection{Historical wind power productions}
\begin{figure*}[th]
\centerline{%
    \begin{subfigure}{0.42\textwidth}
        \centering
        \includegraphics[width=\textwidth]{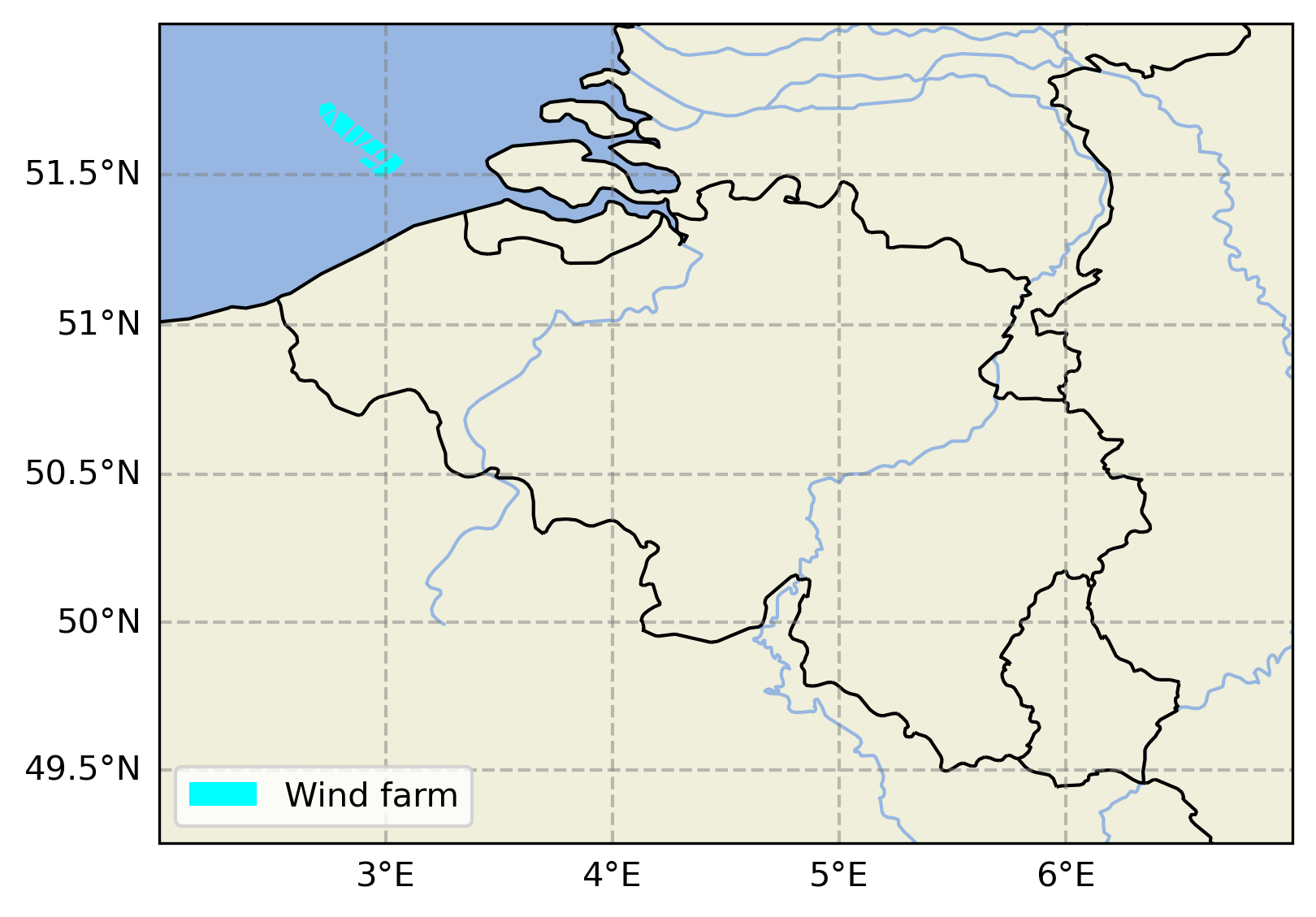} 
    \end{subfigure}
    \hfill
    \begin{subfigure}{0.555\textwidth}
        \centering
        \includegraphics[width=\textwidth]{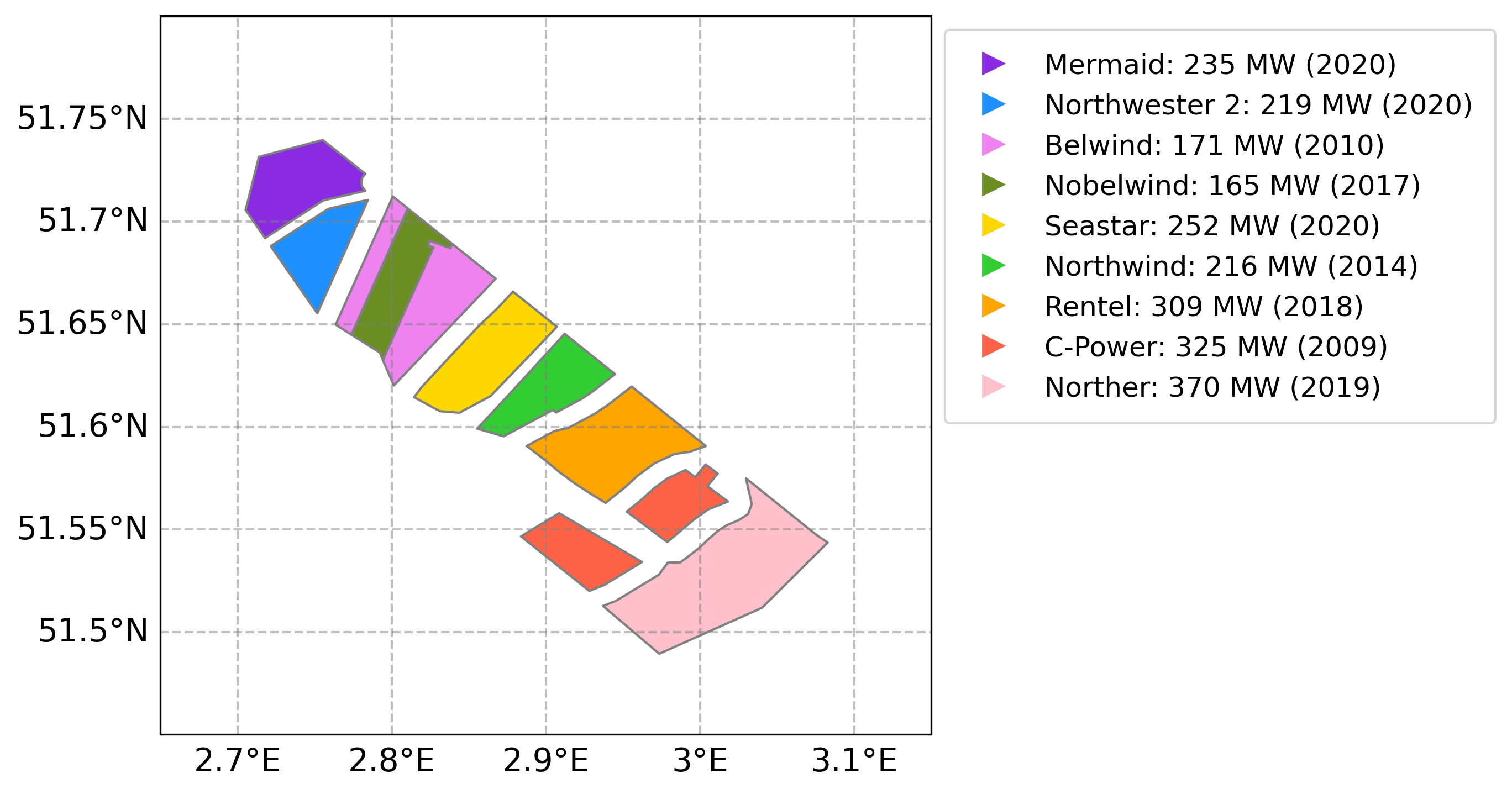} 
    \end{subfigure}%
}
\caption{The location of BOZ wind farms and installed capacity of each wind farm.}
\label{fig:BOZ_map}
\end{figure*}

The BOZ currently comprises 11 operational wind farms with a combined installed capacity of 2262 MW (figure \ref{fig:BOZ_map}). The historical offshore wind power production data for BOZ used in this study is sourced from Elia's Open Data Platform (available at \url{https://opendata.elia.be/explore/dataset/ods031/information/}). This dataset offers detailed 15-minute interval records of aggregated wind power output from BOZ.

In this work, we focus on the period for power data records of years 2022 and 2023. The dataset also includes a ``Decremental Bid Indicator'', which indicates instances of manual power reduction to reach bidding transactions. These date-times are removed from our dataset. While the platform also provides a ``Most recent forecast'' archive, the wind power predictions discussed in this paper are derived from the RMI's NWP wind forecasts with specific power conversion methods, which differ from Elia's forecast archive. 

\subsection{Wind power forecasts based on NWP models}
\subsubsection{NWP model description}
The ALARO-4km model (ALO4) is the operational NWP system at RMI, which is developed within the ALADIN consortium and tailored for regional-scale weather forecasting \cite{termonia2018aladin}. It is based on a code that is shared with the global model Integrated Forecast System (IFS) of the ECMWF and the ARPEGE model of Météo-France, and its configurations are particularly coupled to the ARPEGE model. It operates at a horizontal resolution of 4 km and outputs 15-minute averaged wind speeds up to a forecast lead time of 60 hours. The wind speed forecasts are archived at multiple levels and interpolated to the turbine hub height.

The operational ALARO-4km model does not account for wind farm wake interactions between turbines. To address this limitation, Fitch's method of Wind Farm Parameterization (WFP) has been incorporated into the ALARO-4km model \cite{fitch2012local, dieter2025improving}. The WFP-enhanced model, denoted as ALO4-WFP, generates wind forecasts that account for intra-windfarm wake losses and flow disturbances. 

In addition, the ECMWF High Resolution (HRES) model is an important reference for operational weather forecasts. HRES provides hourly 100m wind speed forecast with a spatial resolution of 0.1° (\textasciitilde 9 km). These forecasts offer a valuable comparison for assessing the accuracy of the ALARO-4km models.

For all NWP models, we use the 00 UTC forecast run and select the nearest grid points to the middle latitude and longitude of each wind farm.

\subsubsection{Power conversion methods}
The operational wind power prediction at RMI uses ALO4 wind speed forecasts at turbine height. The Power Curve (PC) converts wind speed into power by applying the specific power curve function \cite{carrillo2013review}:
$$ P = \frac{1}{2}\rho A C_p  |V|^3$$
where $V$ is the wind speed, $\rho$ is the air density, $A$ is the rotor area and $C_p$ is the power coefficient, which are turbine technical specifications related to the blade design, the tip angle, etc. The power output of each wind farm is calculated with an overall correction factor per wind farm, based on historical power data, and the total BOZ power is obtained by summing the outputs of all wind farms.

\begin{figure}[ht]
    \centering
    \begin{subfigure}{0.45\textwidth}
        \centering
        \includegraphics[width=\textwidth]{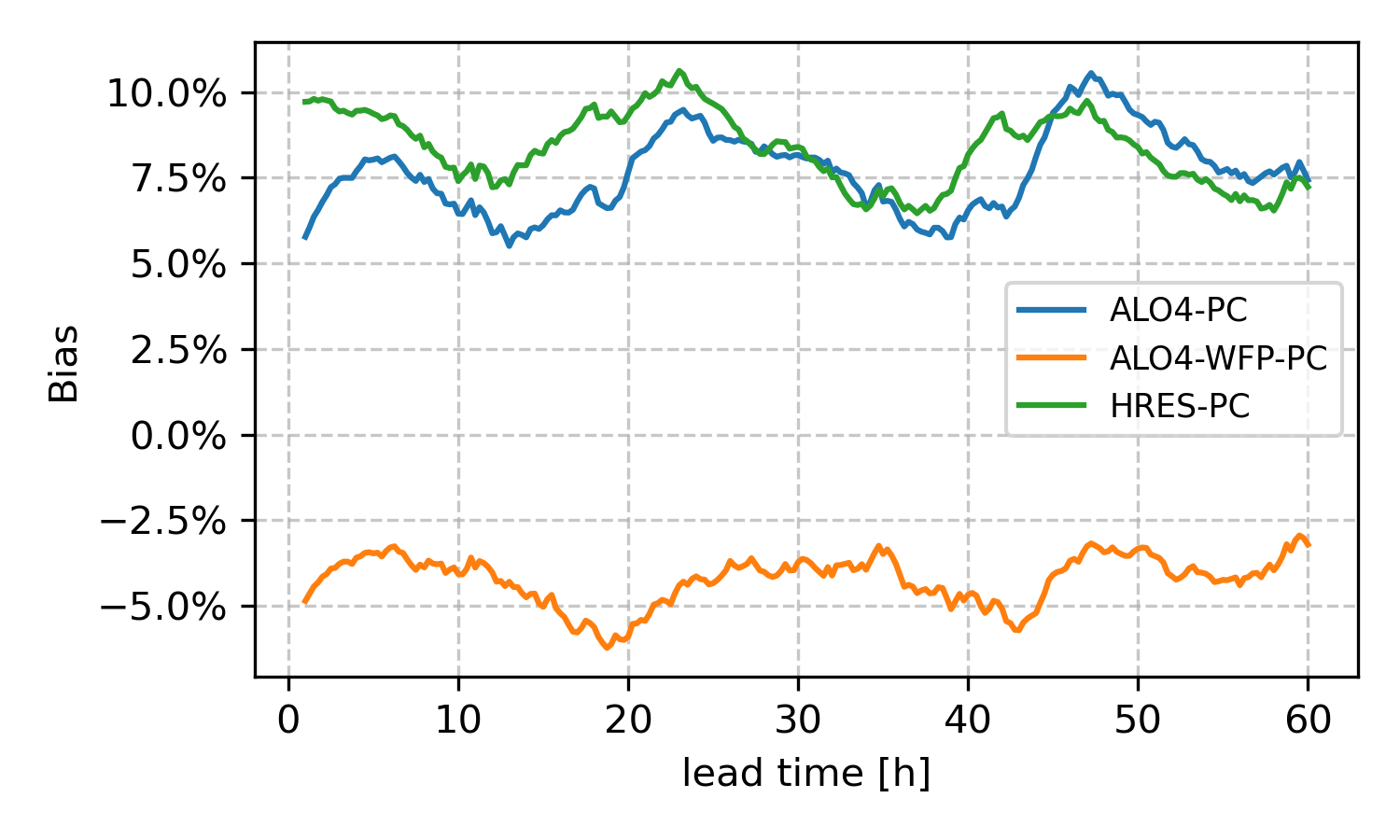} 
    \end{subfigure}
    \hfill
    \begin{subfigure}{0.45\textwidth}
        \centering
        \includegraphics[width=\textwidth]{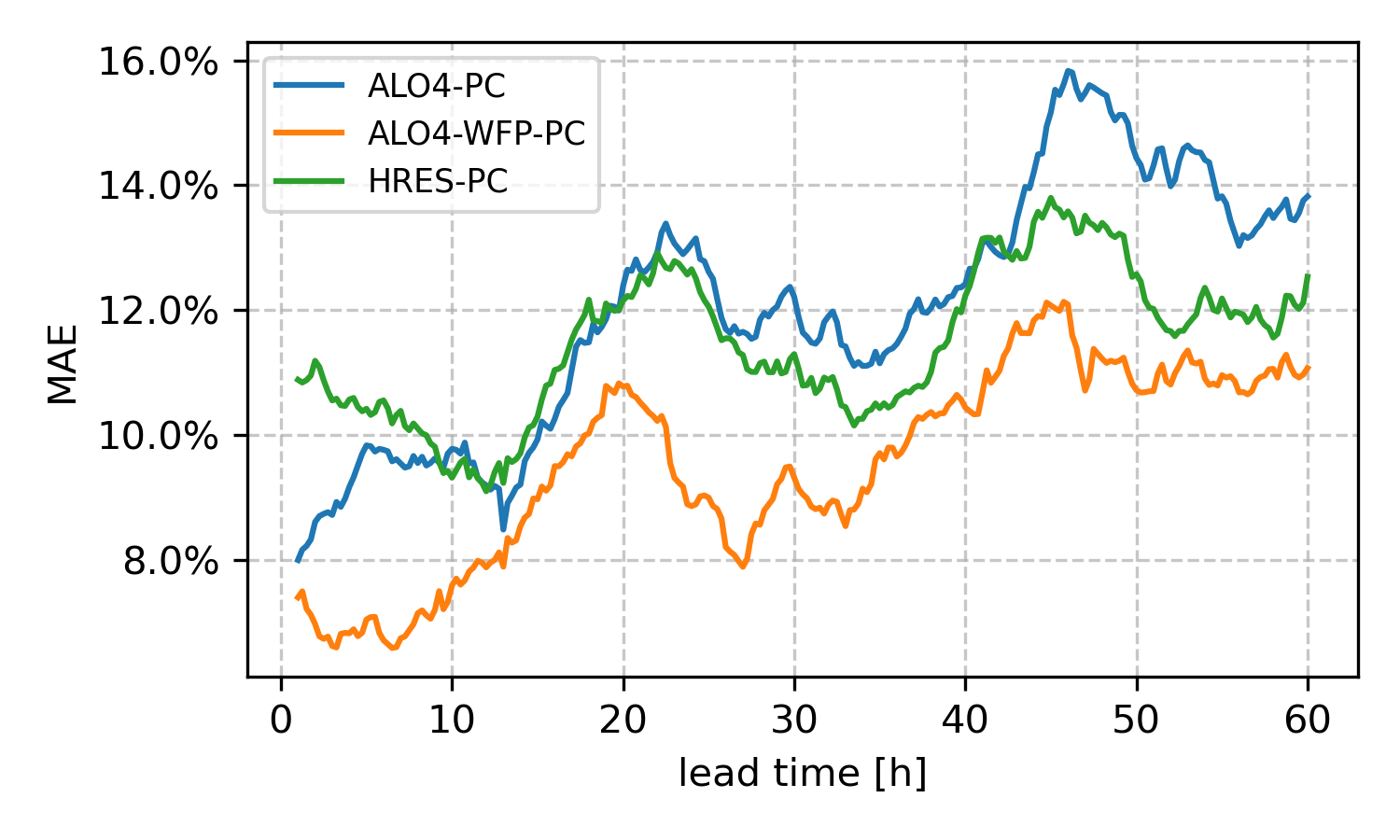} 
    \end{subfigure}
    \caption{PC model prediction bias (top) and MAE (bottom) of BOZ aggregated wind power predictions against power production in percentage of installed capacity, verified over the years 2022 and 2023.} 
    \label{fig:WP_PC_bias_mae}
\end{figure}

When applying the PC to the aforementioned NWP models for wind speed-to-power conversion, the performance differences between models can be verified (figure \ref{fig:WP_PC_bias_mae}). For the operational ALO4 model, wind power forecasts exhibit a clear positive bias, likely because the model does not account for wake effects within the wind farm, leading to an overestimation of wind speed and, consequently, a positive bias in power output. In contrast, the ALO4-WFP-PC model consistently exhibits a negative bias. This may be attributed to two factors: first, the wake effect estimation in the WFP module may be overly strong; second, the idealized power curve conversion is not well-fitted to the WFP wind forecasts. Despite this bias, the ALO4-WFP-PC model achieves a reduction in Mean Absolute Error (MAE) across all lead times, indicating that the incorporation of WFP improves the overall accuracy of power forecasts.

By comparison, the HRES-PC exhibits the largest bias at most lead times and has the highest MAE within the first 10 lead hours. Between 10–24h, its MAE is comparable to that of ALO4-PC, while at longer lead times it falls below ALO4-PC but remains higher than ALO4-WFP-PC. This can be explained by three main reasons: first, HRES does not include any wake effect correction, resulting in wind speed overestimation; second, its coarser spatial resolution compared to ALO4 means that multiple wind farms share the same grid cell forecast, reducing the representativeness and accuracy of wind speed inputs used for power conversion; third, HRES-PC uses wind speed at 100m rather than at turbine height for power conversion.

Considering that HRES-PC does not use wind speed predictions at the same height as ALO4-PC and ALO4-WFP-PC, this is not necessarily a fair comparison. Specifically, HRES-PC utilizes wind speeds at a standard 100-meter height, whereas the ALO4-based models use interpolated turbine-height wind speeds, creating a fundamental difference in the input variables. Furthermore, the grid coordinates of HRES are also different from those of ALO4. Therefore, the HRES serves primarily as a reference model. To specifically evaluate the performance of the WFP module on power prediction, the subsequent discussion will be limited exclusively to the ALO4-based NWP models. 

\subsubsection{Machine learning methods}

As mentioned in the previous section, although ALO4-WFP-PC achieves a lower MAE compared to ALO4-PC, it exhibits a negative bias, possibly in part because the idealized power curve conversion is not fully compatible with the wind speed forecasts from the WFP. This issue motivates exploring alternative approaches for the power conversion. Nowadays, Machine Learning (ML) is popular in handling the complex dependency between multiple variables. Here we use the ALO4-WFP meteorological forecast variables as input to train ML models to investigate the predictive skill of ML-based power forecasting in comparison to classical PC conversion. The models are trained on data of the year 2022 and validated on 2023, ensuring that both the training and validation periods cover a full annual cycle. To provide a robust comparison, we include two ML models:

\begin{itemize}
    \item \textbf{Neural Network} 
    
    A Neural Network (NN) is designed with inputs of wind speed, wind direction, and lead time, which allows models to learn potential wake effect differences by wind direction and correct lead time-dependent biases \cite{dieter2025improving}. The NN model is implemented using the Keras framework and follows a multi-layer perceptron architecture. The input layer uses three variables of wind speed, direction, and lead time. Then the model contains three fully connected hidden layers with 64 nodes each and ReLU activation functions. The output layer consists of a single node that predicts wind power at the corresponding lead time. The model is trained using the Adam optimizer with a learning rate of 0.001, and optimized with the MAE as the loss function. Training is conducted with a batch size of 256 and employs early stopping to prevent overfitting.

    \item \textbf{XGBoost} 
    
    An XGBoost model (XGB), which is an efficient implementation of gradient-boosted decision trees, is also implemented with the same input variables as the NN model. The XGBRegressor class from the xgboost Python library is applied with default hyperparameter settings, including 100 estimators, a learning rate of 0.1, and a maximum tree depth of 3 \cite{chen2016xgboost}. Multiple hyperparameter configurations are also tested, but the results show little variation, indicating that the default settings provide sufficiently robust performance for this application. The XGB model is also optimized with the MAE loss function. 

\end{itemize}

\begin{figure}[h]
    \centering
    \begin{subfigure}{0.45\textwidth}
        \centering
        \includegraphics[width=\textwidth]{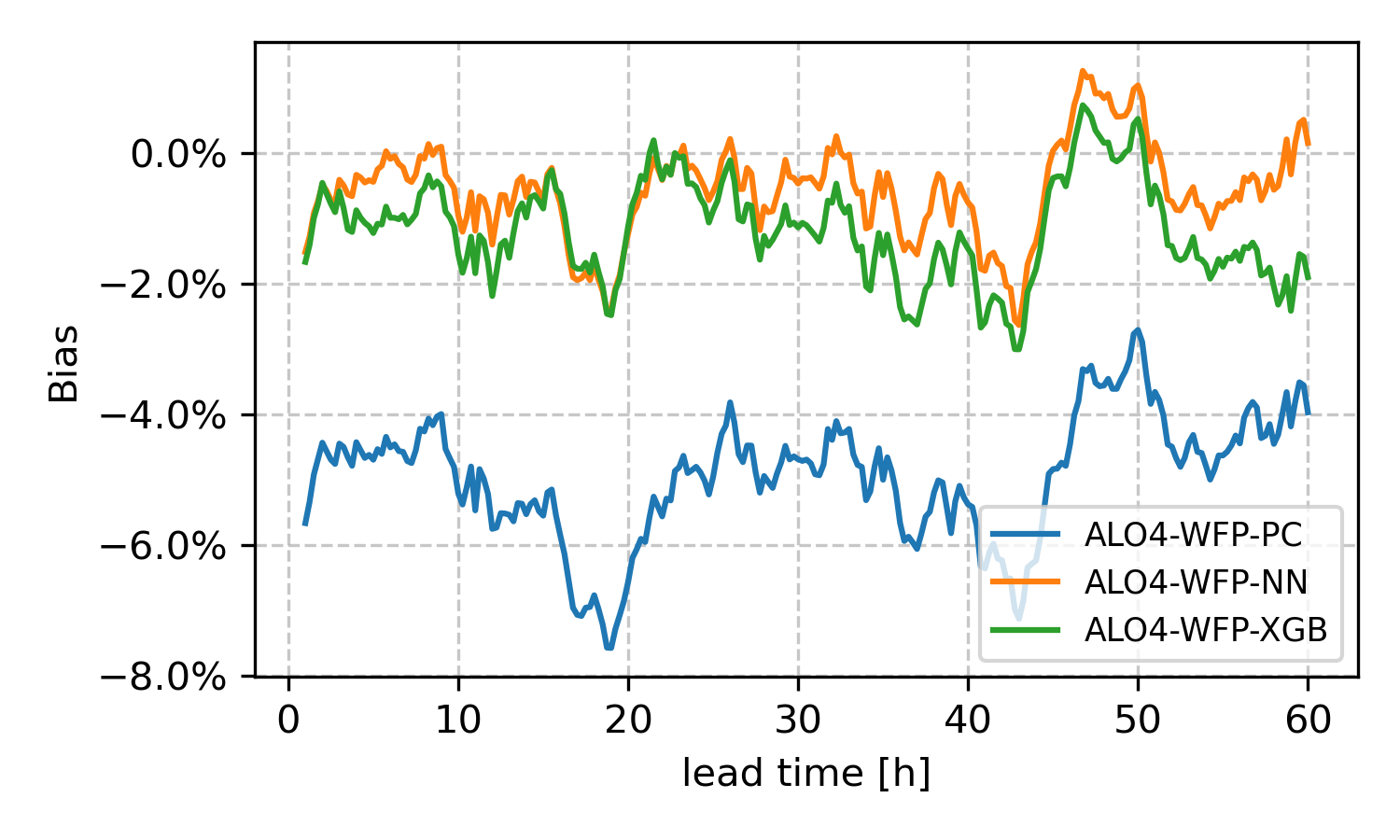} 
    \end{subfigure}
    \hfill
    \begin{subfigure}{0.45\textwidth}
        \centering
        \includegraphics[width=\textwidth]{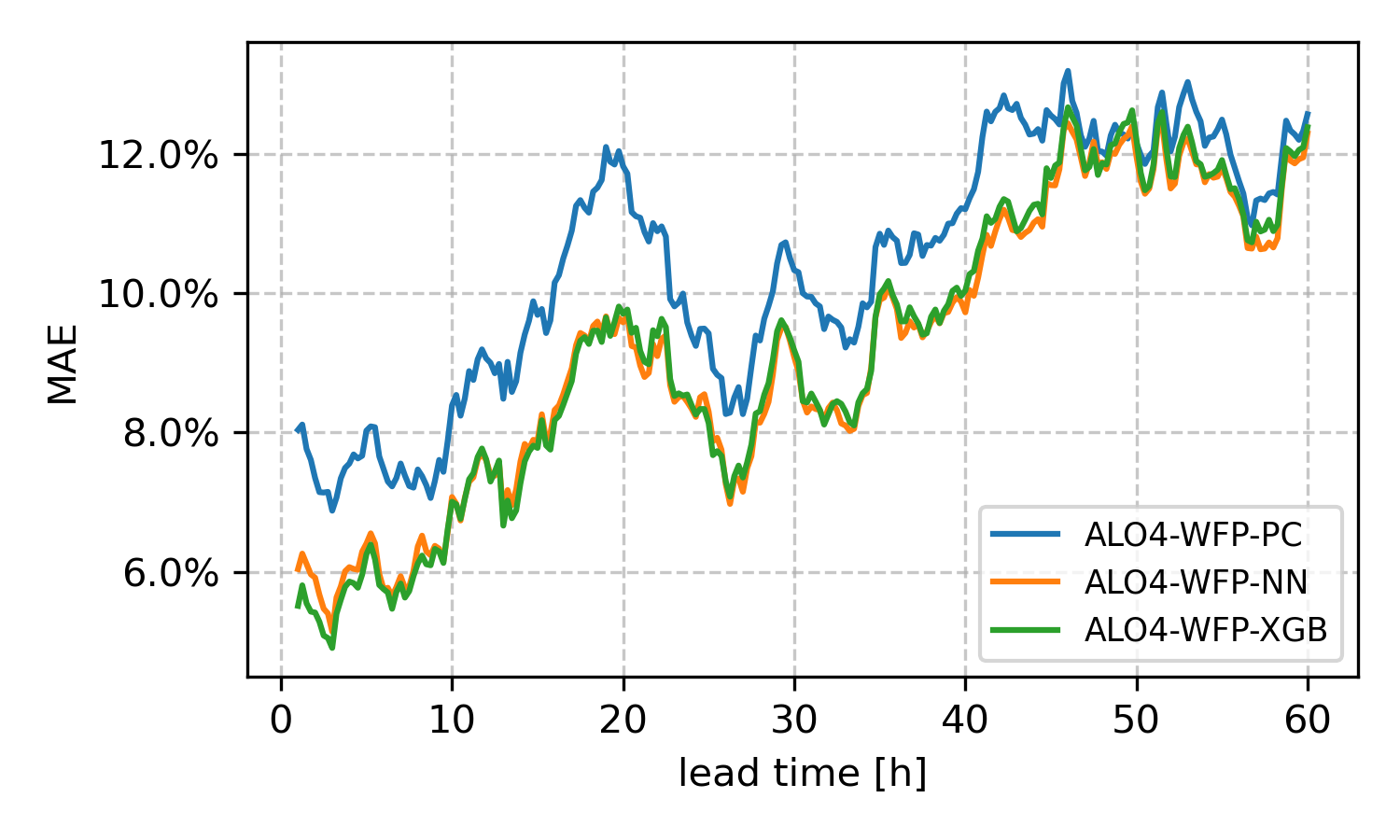} 
    \end{subfigure}
    \caption{ML model bias (top) and MAE (bottom) of BOZ power predictions in percentage of installed capacity, verified over the year 2023.} 
    \label{fig:WP_ML_bias_mae}
\end{figure}

The primary advantage of the ML models is their ability to effectively correct the bias (figure \ref{fig:WP_ML_bias_mae}). Compared to ALO4-WFP-PC, which exhibits an average bias of approximately -5\% over a 60-hour lead time, all ML models reduce the mean bias to -1\%. They also achieve a lower average MAE, with the improvement particularly notable at shorter lead times. After 48 lead hours, the MAE of the ML models approaches that of the PC results.

\section{Power ramping and ramping event predictability}

\subsection{Definition of power ramping and ramping event}
Wind power ramping typically refers to a significant and rapid wind power fluctuation in wind farms. Key characteristics of a ramp include its magnitude (the power change $\Delta P$), and duration (the time window over $\Delta t$). Additionally, the timing (often defined as either the starting or central time of the event) and the direction (whether it represents an increase or decrease in power) are essential parameters \cite{gallego2015review, ferreira2011survey}. The power ramping values can be computed as the power difference within the analysis window between two selected time points. Positive and negative values represent up and down ramps respectively. In this study, we only consider cases where the daily maximum wind speed is below 20 m/s to exclude the effects of extreme wind conditions, such as storms, which could cause cut-outs \cite{elia2018offshore}. This filtering results in an exclusion of 14\% of the dates from the verification period. Samples containing decremental bids in the dataset are also removed. These constraints allow us to focus specifically on power ramping driven by meteorological variability rather than the mechanical limitations of wind turbines or operational energy dispatching reasons. The following analysis specifically focuses on 15-minute and 1-hour ramping, as Elia has expressed an interest in short-term, rapid power ramps, which are critical for grid stability and operational decision-making \cite{elia2026adequacy}. 

Ramping events are defined as instances where the power difference between the starting time and ending time exceeds a specified threshold. To illustrate the frequency of ramping events under different thresholds, we present the number of days per year with at least one 15-minute and 1-hour ramping event occurring (table \ref{tab:ramping_event_count}).

\begin{table*}[ht]
\centering
\begin{tabular}{cccccccccc}
\hline
                                            &                           & -50\% & -30\% & -15\% & -10\% & 10\% & 15\% & 30\% & 50\% \\ \hline
\multicolumn{1}{l|}{\multirow{2}{*}{15min}} & \multicolumn{1}{c|}{2022} & 1     & 7     & 50    & 105   & 117  & 53   & 5    & 1    \\ \cline{2-10} 
\multicolumn{1}{l|}{}                       & \multicolumn{1}{c|}{2023} & 1     & 7     & 53    & 110   & 139  & 76   & 6    & 1    \\ \hline
\multicolumn{1}{c|}{\multirow{2}{*}{1h}}    & \multicolumn{1}{c|}{2022} & 12    & 44    & 167   & 225   & 222  & 175  & 66   & 10   \\ \cline{2-10} 
\multicolumn{1}{c|}{}                       & \multicolumn{1}{c|}{2023} & 13    & 56    & 183   & 240   & 233  & 179  & 70   & 19   \\ \hline
\end{tabular}
\caption{Number of days per year of at least one ramping event occurring in the BOZ, categorized by multiple ramping thresholds in percentage of BOZ total capacity. Dates characterized by a daily maximum wind speed exceeding 20 m/s and the date times with decremental bidding transactions are excluded.}
\label{tab:ramping_event_count}
\end{table*}

\subsection{Ramping prediction assessment for power models}

The previous section demonstrates that ML models can achieve near-zero bias and lower MAE in power prediction compared to traditional power curve conversions. However, early studies have emphasized that MAE fails to adequately capture the temporal dynamics and amplitude-specific features that characterize ramping analysis \cite{vallance2017towards}. Such metrics provide a broad overview of overall prediction accuracy but may obscure nuances critical for capturing rapid changes in wind power output. 

\begin{figure}[h]
    \centering
    \begin{subfigure}{0.45\textwidth}
        \centering
        \includegraphics[width=\textwidth]{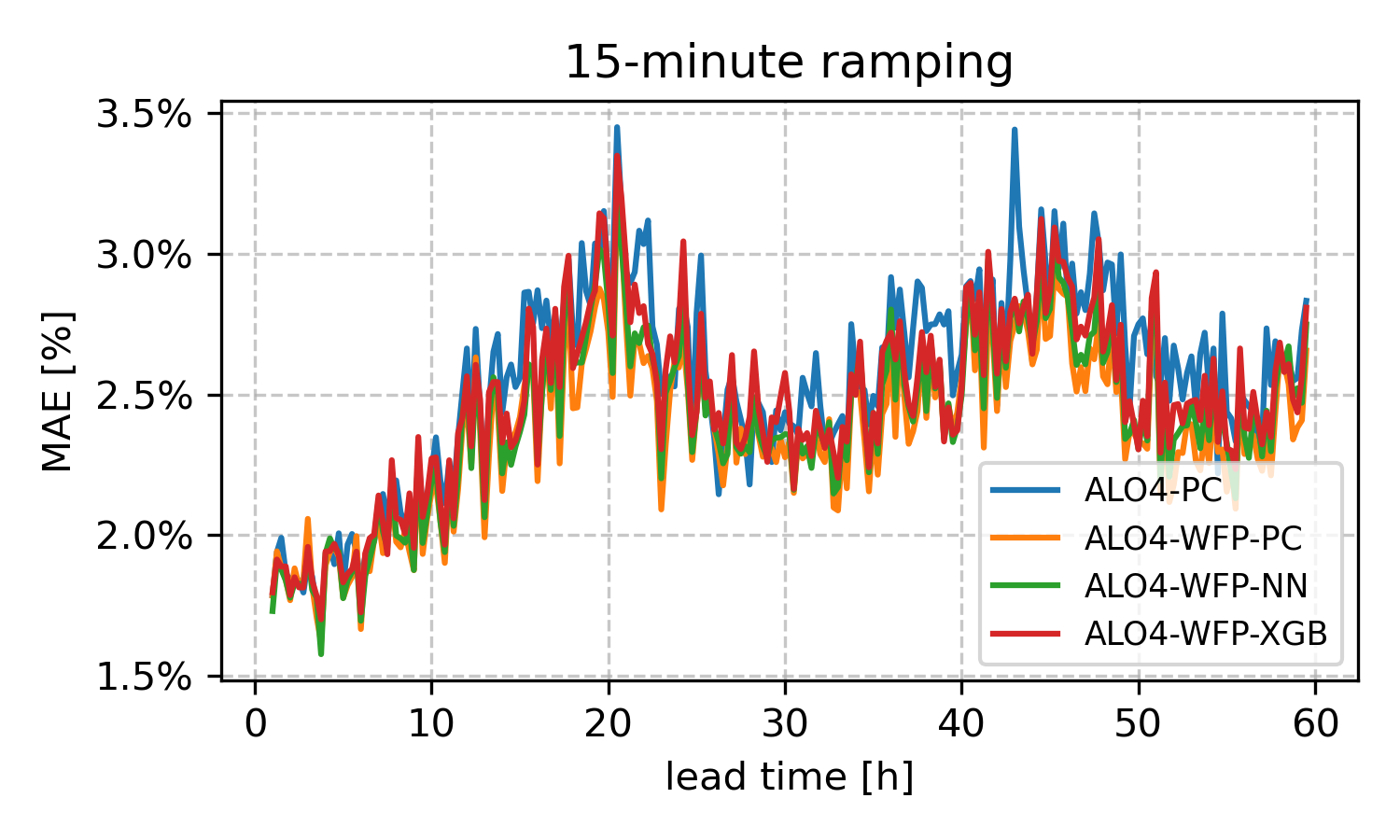} 
    \end{subfigure}
    \hfill
    \begin{subfigure}{0.45\textwidth}
        \centering
        \includegraphics[width=\textwidth]{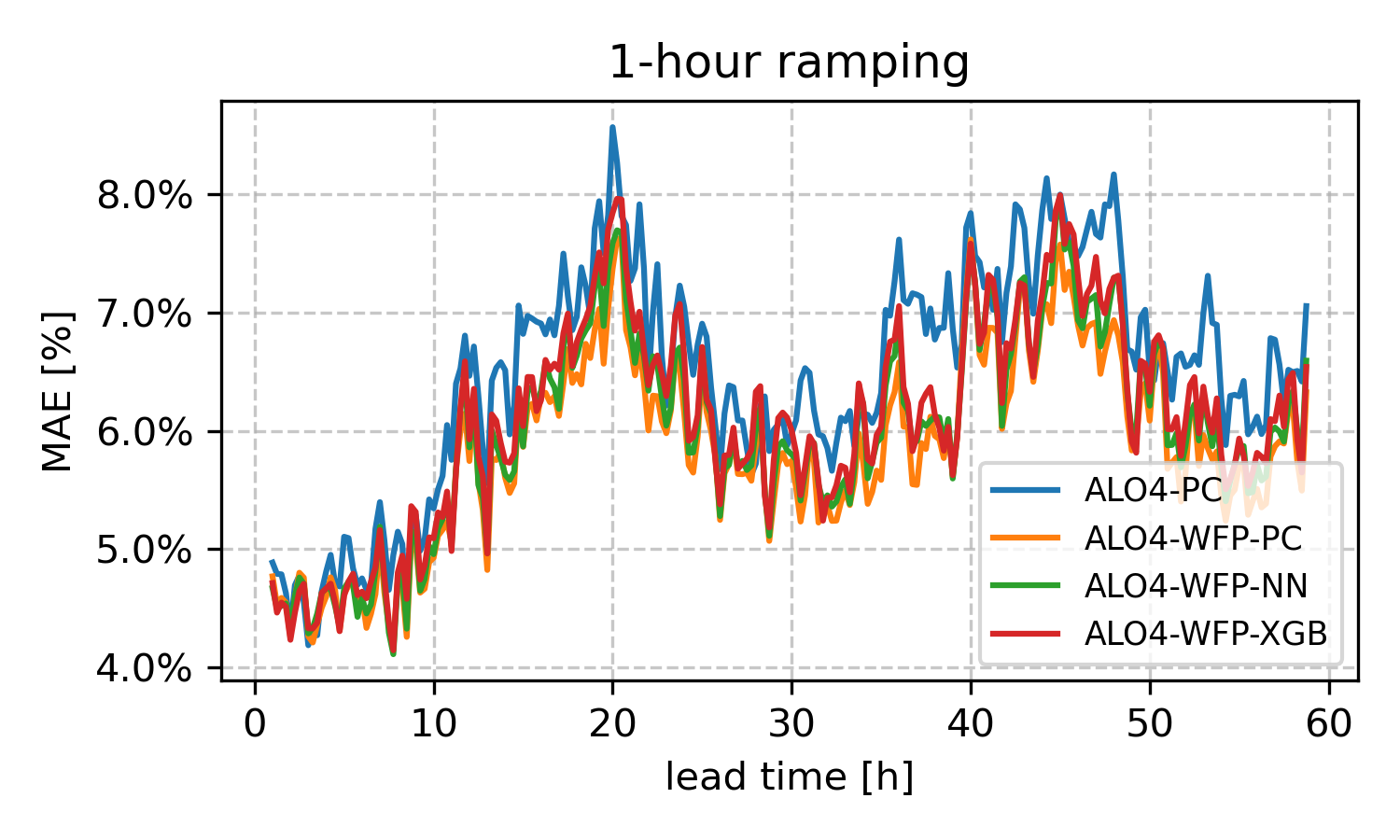} 
    \end{subfigure}
    \caption{15-minute ramping MAE (top) and 1-hour ramping MAE (bottom) of BOZ power ramping in percentage to the total capacity. The verification period is year 2023.}
    \label{fig:WP_ramp_mae}
\end{figure}

The models’ ramping time series are obtained from power differences over fixed intervals, where each lead-time ramping value represents the power change over the subsequent period. The errors of these predicted ramps are then validated against the corresponding observations with the ramping MAE. The results reveal that a lower power prediction MAE does not guarantee a better representation of ramping prediction (figure \ref{fig:WP_ramp_mae}). While the ML models demonstrate significantly lower MAE in predicting total wind power, their ramping MAE is not consistently lower than ALO4-WFP-PC. Moreover, while the MAE of wind power prediction generally exhibits an increasing trend with longer lead times, this pattern is not as evident in ramping MAE. Instead, the ramping MAE as a function of lead time displays a more pronounced diurnal pattern, characterized by higher errors during daytime hours and lower errors at night. 

Table \ref{tab:ramping_MAE} reports the ramping MAE averaged over all lead times, as well as separately for intraday forecasts (lead times $\leq$ 24h) and day-ahead forecasts (24-48h). Although ML power forecasts achieve lower average errors, their corresponding ramping values don't necessarily translate to lower MAE, demonstrating that relying solely on MAE verification is insufficient to understand the actual performance of these models in predicting ramping events.

\begin{table*}[ht]
\centering
\begin{tabular}{c|cc|cc|cc|cc}
\hline
\multirow{2}{*}{} & \multicolumn{2}{c|}{ALO4-PC}              & \multicolumn{2}{c|}{ALO4-WFP-PC}          & \multicolumn{2}{c|}{ALO4-WFP-NN}          & \multicolumn{2}{c}{ALO4-WFP-XGB}          \\ \cline{2-9} 
                  & \multicolumn{2}{c|}{all-lead-time}        & \multicolumn{2}{c|}{all-lead-time}        & \multicolumn{2}{c|}{all-lead-time}        & \multicolumn{2}{c}{all-lead-time}         \\ \hline
15min             & \multicolumn{2}{c|}{2.6\%}                & \multicolumn{2}{c|}{2.4\%}                & \multicolumn{2}{c|}{2.4\%}                & \multicolumn{2}{c}{2.5\%}                 \\ \hline
1h                & \multicolumn{2}{c|}{6.5\%}                & \multicolumn{2}{c|}{5.9\%}                & \multicolumn{2}{c|}{6.0\%}                & \multicolumn{2}{c}{6.1\%}                 \\ \hline
                  & \multicolumn{1}{c|}{intraday} & day-ahead & \multicolumn{1}{c|}{intraday} & day-ahead & \multicolumn{1}{c|}{intraday} & day-ahead & \multicolumn{1}{c|}{intraday} & day-ahead \\ \hline
15min             & \multicolumn{1}{c|}{2.5\%}    & 2.7\%     & \multicolumn{1}{c|}{2.3\%}    & 2.6\%     & \multicolumn{1}{c|}{2.4\%}    & 2.6\%     & \multicolumn{1}{c|}{2.4\%}    & 2.6\%     \\ \hline
1h                & \multicolumn{1}{c|}{6.2\%}    & 7.0\%     & \multicolumn{1}{c|}{5.7\%}    & 6.2\%     & \multicolumn{1}{c|}{5.9\%}    & 6.4\%     & \multicolumn{1}{c|}{5.9\%}    & 6.5\%     \\ \hline
\end{tabular}
\caption{MAE of model power ramping predictions in percentage of BOZ total installed capacity. Values are presented for the averaged values across all lead times, intraday and day-ahead. }
\label{tab:ramping_MAE}
\end{table*}

Wind power ramping analysis fundamentally concerns power variability in time series. In signal processing, such variability can be effectively characterized by the Power Spectral Density (PSD), which quantifies how the power of a time series is distributed across different frequency components \cite{lee2012analyzing}. The PSD of wind power time series quantifies the extent to which, compared to the observation, different models capture variability across multiple timescales, from low-frequency to high-frequency (ramping) behavior. Welch’s method is applied for PSD analysis. Each time series of observations and predictions is divided into 128 segments, and a sampling frequency of 96 is applied, corresponding to the 15-minute resolution of the data (i.e., 96 time steps per day). The frequencies are represented in units of days$^{-1}$, in accordance with common scales in atmospheric spectral analysis \cite{larsen2012recipes}. The PSD analysis is performed individually on each predicted time series, and the resulting spectra are then averaged across all samples to obtain the mean PSD values at each frequency. The frequency of 12 day$^{-1}$ means an up-down cycle every two hours, which corresponds to two hourly ramping processes. 

\begin{figure}[htbp]
    \centering
    \includegraphics[width=0.45\textwidth]{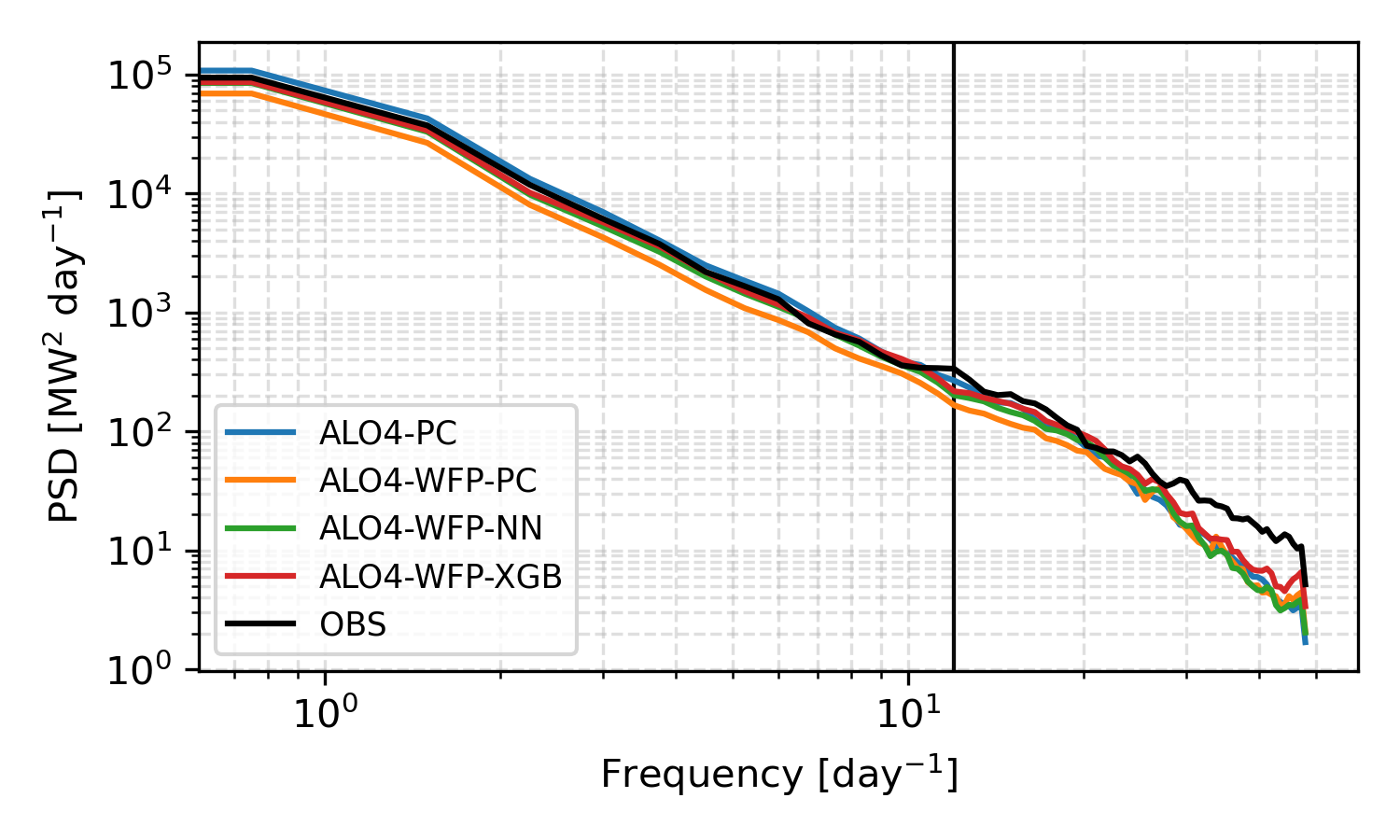}
    \caption{PSD of wind power time series of observation and model predictions. The vertical line at 12 day$^{-1}$ represents the high-frequency ramping events.}
    \label{fig:PSD}
\end{figure}

In the low-frequency range ($<$ 12 day$^{-1}$), the PSD of the ALO4-PC model is slightly higher than that of the observations, but it turns to an underestimation at higher frequencies ($\geq$ 12 day$^{-1}$)  (figure \ref{fig:PSD}). This indicates its overestimation of ramping intensity at longer timescales, while an underestimation at hourly and sub-hourly scales. In contrast, the ALO4-WFP-PC model exhibits consistently lower PSD values across all frequency ranges, suggesting insufficient variability in its predictions. This infers the limitation of representing an entire wind farm with a single grid point from the WFP model: wake effects from individual turbines propagate downstream, and insufficient spatial resolution dampens intra-farm variability. Machine learning models partially mitigate this issue, as indicated by PSDs closer to the observations, by better capturing the statistical distribution. Nevertheless, all models tend to underestimate short-time ramping intensity, as their PSDs are lower than the observation at higher frequencies. The PSD pattern indicates that XGB better captures high-frequency variability compared to NN. Therefore, in the subsequent analysis, XGB is selected as the representative ML model to further investigate its skill in ramping event predictions.

These results collectively demonstrate that a lower MAE in power prediction does not guarantee an improved performance in capturing the dynamics of power ramping, and such a metric is not capable of comprehensively understanding the predictability of ramping events. Therefore, ramp-tailored verification schemes are essential to properly evaluate model capability in ramping predictions.

\subsection{Ramping event verification incorporating buffers}

The interpretation of ``large'' and ``rapid'' ramping is not universally standardized. It depends on factors such as wind farm size and the forecast model's ultimate application (e.g., market penalties versus instantaneous demand response). To enhance the ramping event predictability assessments, we have developed a dedicated verification framework that is based on the binary occurrence of events for the contingency table with assignable ramping magnitude and duration parameters, while introducing two additional parameters: time buffer and power buffer. The time buffer allows flexibility in assessing forecast timing errors by accepting events that occur within a certain temporal window around the observed ramps. Specifically, a ramping event is defined as a power change that exceeds a predefined threshold $\theta$ over a fixed time interval $\Delta t$. The sets of observed and predicted ramping time series are defined as:

$$
R_\text{obs} = \left\{ t \in T \,\middle|\, \left|P_\text{obs}(t + \Delta t) - P_\text{obs}(t)\right| \geq \theta \right\},
$$

$$
R_\text{pred} = \left\{ t \in T \,\middle|\, \left|P_\text{pred}(t + \Delta t) - P_\text{pred}(t)\right| \geq \theta \right\},
$$

where $P_\text{obs}(t)$ and $P_\text{pred}(t)$ represent the observed and predicted power at lead time $t$, respectively. A hit is marked if a predicted ramping event at time $t_p \in R_\text{pred}$ is matched by an observed ramping event $t_o \in R_\text{obs}$ in the time interval $|t_p - t_o| \leq \tau$, where $\tau$ is a predefined buffer time. A hit is accordingly expressed as:

$$
\text{Hit} = \left\{ t_o \in R_\text{obs} \,\middle|\, \exists t_p \in R_\text{pred}, |t_o - t_p| \leq \tau \right\}
$$

A miss is defined when an observed ramp exceeds the threshold, but no matching predicted ramp exists within the $\pm \tau$ time buffer:

$$
\text{Miss} = \left\{ t_o \in R_\text{obs} \,\middle|\, \forall t_p \in R_\text{pred}, |t_o - t_p| > \tau \right\}
$$

A false alarm occurs when a forecasted ramp exceeds the threshold but no corresponding observed ramp is found within the $\pm \tau$ time buffer:

$$
\text{False alarm} = \left\{ t_p \in R_\text{pred} \,\middle|\, \forall t_o \in R_\text{obs}, |t_p - t_o| > \tau \right\}
$$

The power buffer tolerates small mismatches in magnitude, enabling a more tolerant assessment of ramping magnitude prediction. For a given ramping threshold $\theta$, and a power buffer proportion $\beta$, the hit cases are defined as:

$$
\text{Hit} =   \left\{ t \in T \,\middle|\, P_\text{obs}(t) \geq \theta \text{ and } P_\text{pred}(t) \geq \theta - \beta \theta \right\}
$$

This approach ensures the ground truth that the ramping events are actually observed, meanwhile verifying whether the predictions fall within a buffer range $\theta - \beta\theta$ to allow small mismatches near the decision boundary.

Accordingly, misses and false alarms are defined as:
$$
\text{Miss} = \left\{ t \in T \,\middle|\, P_\text{obs}(t) \geq \theta \text{ and } P_\text{pred}(t) < \theta - \beta \theta \right\}
$$

$$
\text{False alarm} =  \left\{ t \in T \,\middle|\, P_\text{obs}(t) < \theta \text{ and } P_\text{pred}(t) \geq \theta + \beta \theta \right\}
$$

\begin{figure*}[htbp]
    \centering
    \includegraphics[width=0.6\textwidth]{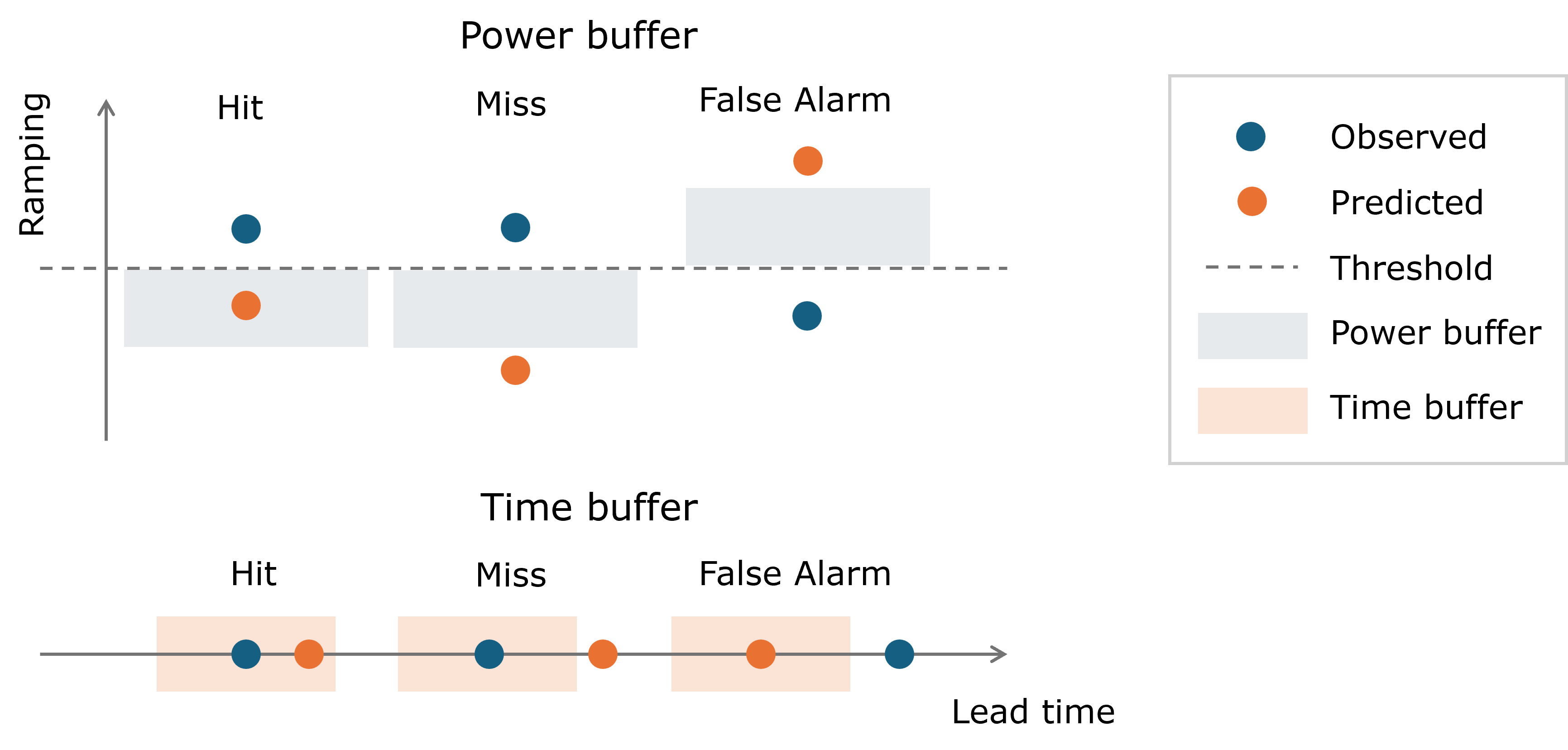}
    \caption{Illustration of power buffer and time buffer verification for ramping events.}
    \label{fig:buffer}
\end{figure*}

These buffer parameters extend the classical binary verification by tolerating minor discrepancies while still crediting correct ramp detections. The time buffer mitigates the impact of a slight forecast lead or lag, whereas the power buffer allows for acceptable deviations in ramp magnitude. Figure \ref{fig:buffer} provides a graphical illustration that intuitively displays how the buffer framework classifies hits, misses, and false alarms. For example, if a forecast predicts a 50\% power drop one hour earlier than an observed 45\% power loss, applying a $\pm$1-hour time buffer and a 10\% power buffer would still classify this forecast as a successful hit. 

With several ramping thresholds, the buffer verification framework classifies forecast outcomes into a standard $2 \times  2$ contingency table: hits (TP), misses (FN), false alarms (FP), and correct negatives (TN). Each combination of magnitude threshold, duration, and buffer settings yields a unique percentage of each metric for each forecasting model. By comparing these percentages across multiple definitions, we can assess how sensitive model performance is to ramp identification criteria and quantify forecast skill under different temporal and magnitude tolerances. We separate the entire forecast range into intraday and day-ahead forecasts and present their respective contingency tables. This separation allows for a comparison of model performance in different forecast periods and avoids the risk of double-counting a single observed ramping event from multiple forecasts extending up to 60 hours. 

\begin{figure*}[h]
    \centering
    \begin{subfigure}{0.49\textwidth}
        \centering
        \includegraphics[width=\textwidth]{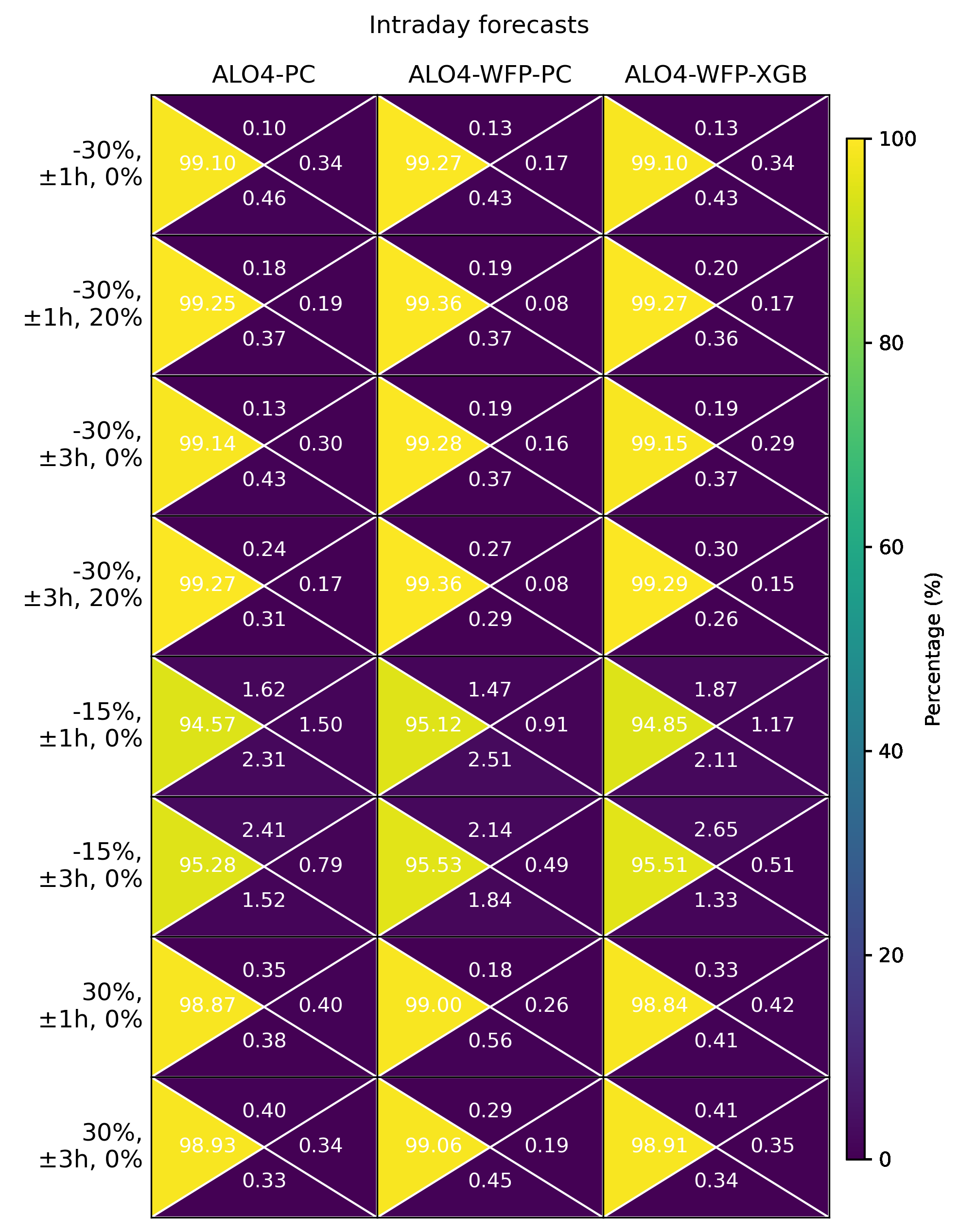} 
    \end{subfigure}
    \hfill
    \begin{subfigure}{0.49\textwidth}
        \centering
        \includegraphics[width=\textwidth]{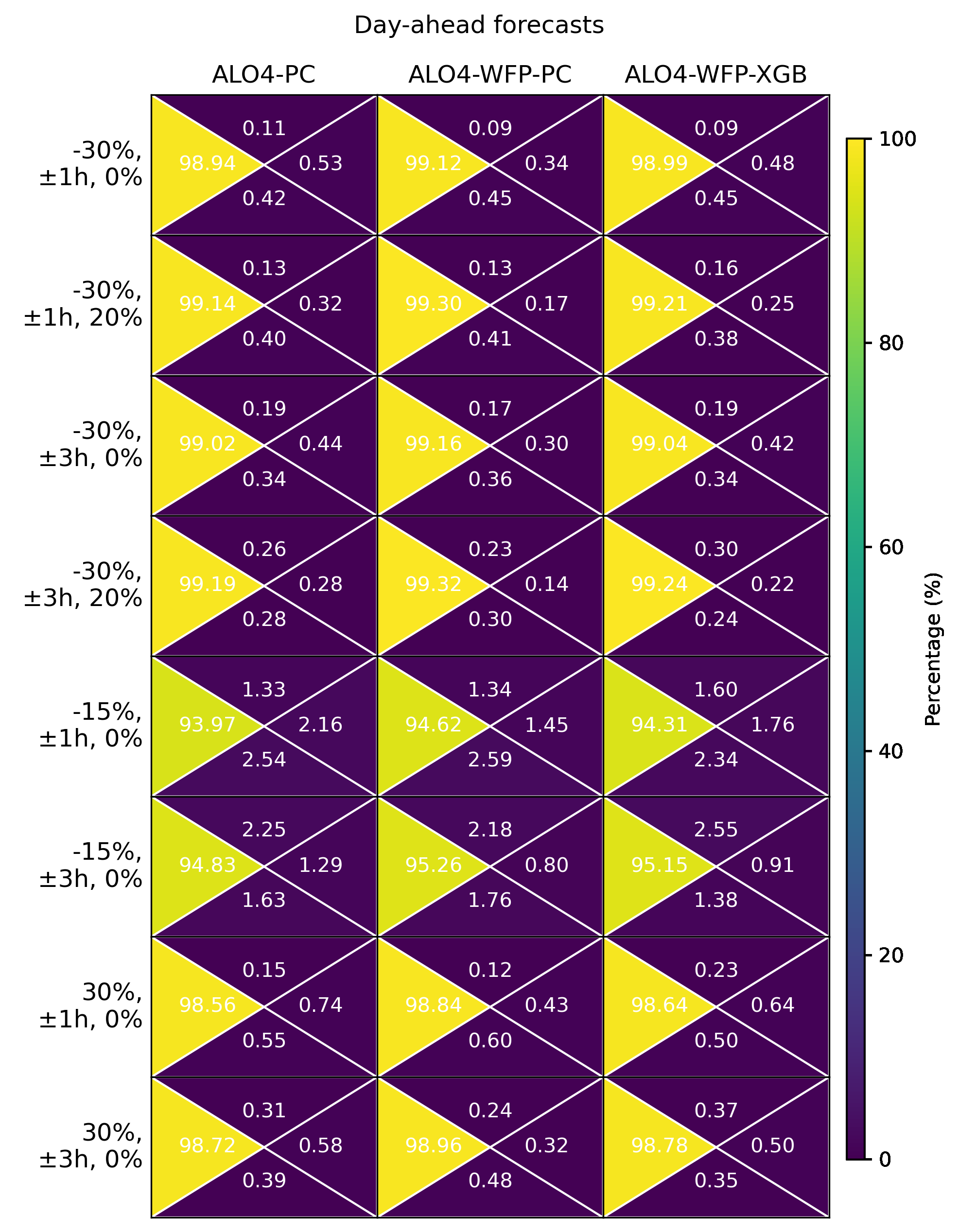} 
    \end{subfigure}
    \caption{The 2 $ \times $ 2 contingency table on the intraday (left) and day-ahead (right) hourly ramping forecasts with various power magnitudes, time buffer, and power buffer in the BOZ. The number in the four triangles in each square suggests the percentages of instances among the four metrics in the contingency table. The triangles on the top, right, bottom, and left are hits, false alarms, misses, and correct negatives, respectively.}
    \label{fig:contingency_tables_1h}
\end{figure*}

Figure \ref{fig:contingency_tables_1h} summarizes the categorical verification results using 2 $\times$ 2 contingency tables for hourly ramping events. These tables illustrate how different ramping intensities are predicted by each model in categorical verification percentages and how the time and power buffer parameters influence the percentages. For all models, the hit percentages of small ramping events are consistently higher than those of large ramping events. Furthermore, increasing both the time buffer and power buffer generally leads to a higher percentage of hit predictions. The hits of intraday forecasts are overall higher than those of day-ahead forecasts, with fewer misses and false alarms. The analysis in this section will focus specifically on ramping predictability in intraday forecasts. The corresponding day-ahead results are provided in the Appendix for reference.

Based on these categorical outcomes, each model’s ramp forecasting skill is quantified by the following verification scores: 
\begin{itemize}
    \item \textbf{Critical Success Index (CSI)}: An overall measure of forecasting skill for ramping events, accounting for correct predictions, misses, and false alarms.
    \[CSI = \frac{TP}{TP + FN + FP}\]

    \item \textbf{Probability of Detection (POD)}: The fraction of actual ramping events successfully captured by the forecast.
    \[POD = \frac{TP}{TP + FN}\]

    \item \textbf{Success Ratio (SR)}: The fraction of predicted ramping events that actually occurred.
    \[SR = \frac{TP}{TP + FP}\]
\end{itemize}

\begin{figure*}[htbp]
    \centering
    \begin{subfigure}{0.9\linewidth} 
        \centering
        \includegraphics[width=\linewidth]{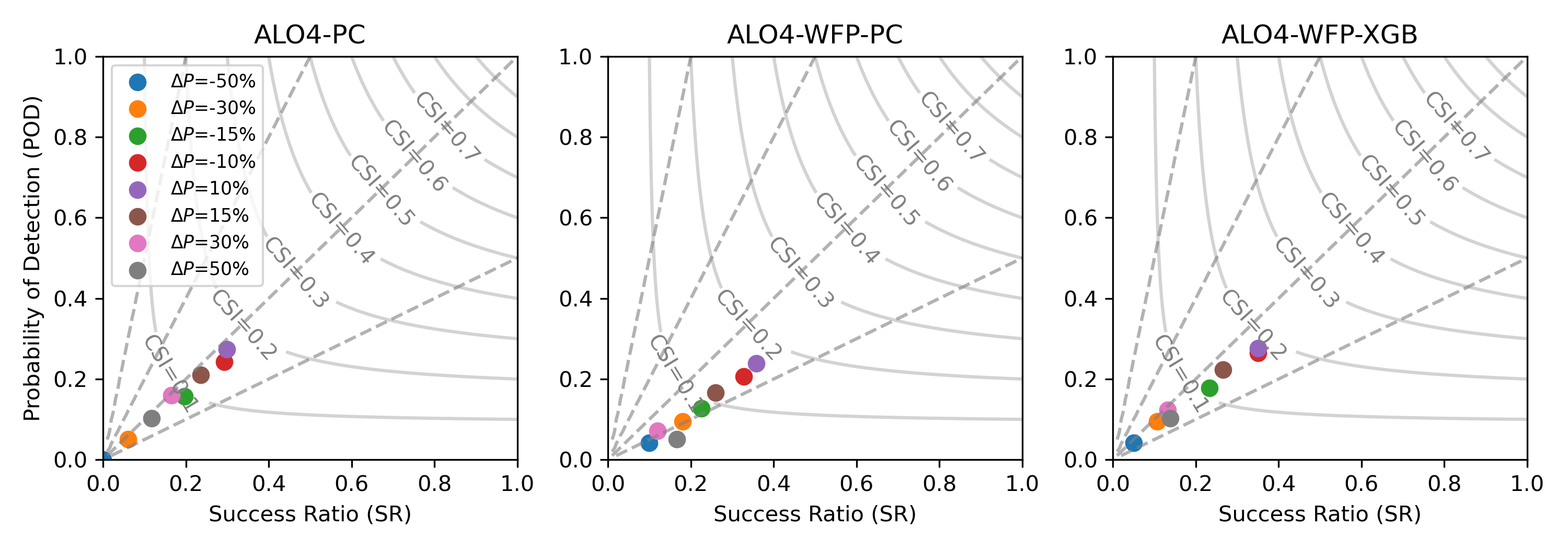} 
    \end{subfigure}

    \begin{subfigure}{0.9\linewidth} 
        \centering
        \includegraphics[width=\linewidth]{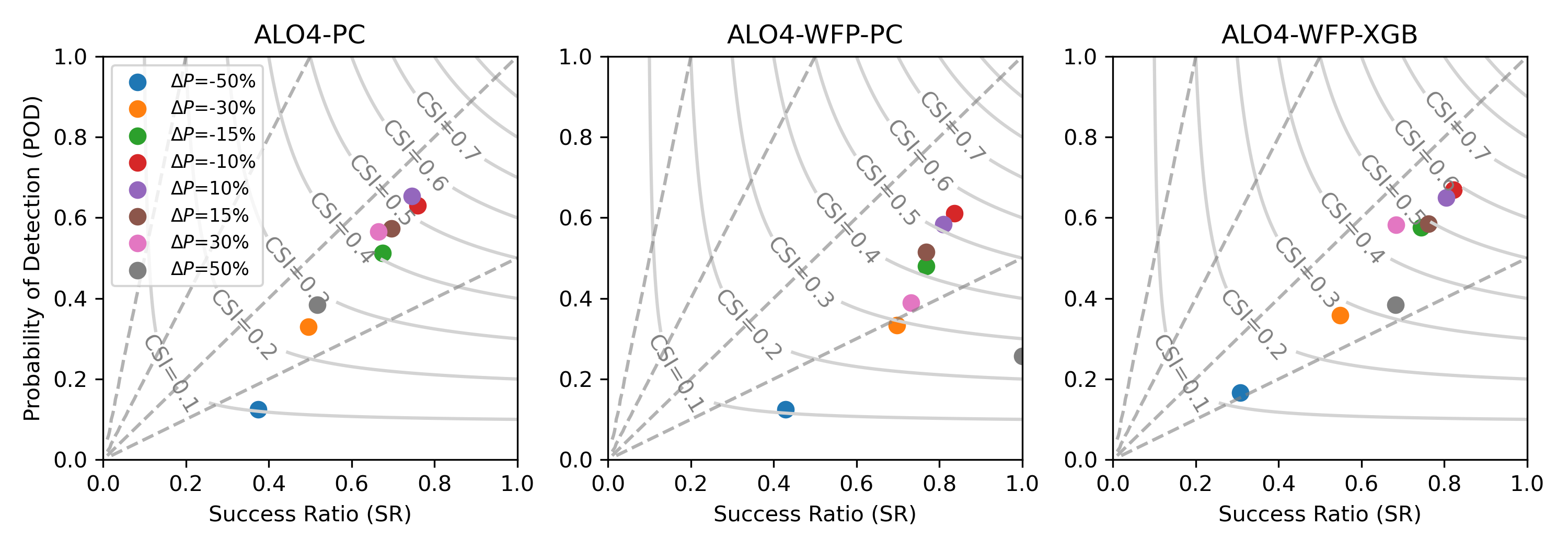} 
    \end{subfigure}

    \caption{Performance diagrams of BOZ ramping event intraday predictions of power models for various hourly ramping thresholds. The figures on the top row use no time buffer and power buffer, and those on the bottom use a 1-hour time buffer and 20\% power buffer.}
    \label{fig:performance_diagram_intraday}
\end{figure*}

These metrics are visualized in performance diagrams \cite{roebber2009visualizing} to compare forecasting performance across different models. Here, we present two sets of hourly ramping event verification results. The first set uses a strict criterion without any buffer, and the second incorporates a 1-hour time buffer and a 20\% power buffer. This comparison serves two purposes: first, to illustrate the impact of buffer settings on verification metrics; and second, to evaluate the predictability of ramping events of varying intensities by each model (figure \ref{fig:performance_diagram_intraday}). 

For strict verification criteria, these models have very low scores for ramping events of all intensities, but once a buffer is allowed, these scores improve significantly and convey more valuable information. This indicates that the power prediction models are generally capable of capturing the direction of ramping events, but struggle with precisely predicting their timing and magnitude. The diagrams show that small ramping events have higher scores than large ones, reflecting the inherent difficulty in capturing intense ramping events. Note that storm events causing cut-outs, which might be more predictable, are excluded from our ramping analysis. Compared to the operational model ALO4-PC, ALO4-WFP-PC has similar CSI, but it gains a higher SR. Recalling the discussion about figure \ref{fig:WP_PC_bias_mae} and \ref{fig:PSD} that ALO4-WFP-PC exhibits a negative bias and lower PSD. The parameterization of wake-effect-induced wind speed reductions results in a fairly conservative power variability and reduces the prediction of large wind speed changes. While this approach reduces the magnitude of predicted ramping events and leads to fewer false alarms, it comes at the cost of missing more such ramping events (as significantly illustrated by the grey dot of the 50\% ramping case). The ALO4-WFP-XGB model achieves higher CSI for all verified ramping thresholds except -50\%, demonstrating an enhanced ramping event predictability. Furthermore, the CSI scores of up-ramping events are consistently higher than those of down-ramping events. This could be attributed to the possibility that the atmospheric processes responsible for up-ramping events (e.g., cold frontal passages and thunderstorm outflows) are generally more predictable in NWP models than relaxation after cold front or boundary convective instability, which are associated with down ramps \cite{zack2006optimization}. Note that extremely large ramping events (e.g., $\pm$50\% hourly ramping) occur infrequently, meaning their verification relies on a limited number of cases. 

\begin{figure*}[htbp]
    \centering
    \includegraphics[width=0.95\linewidth]{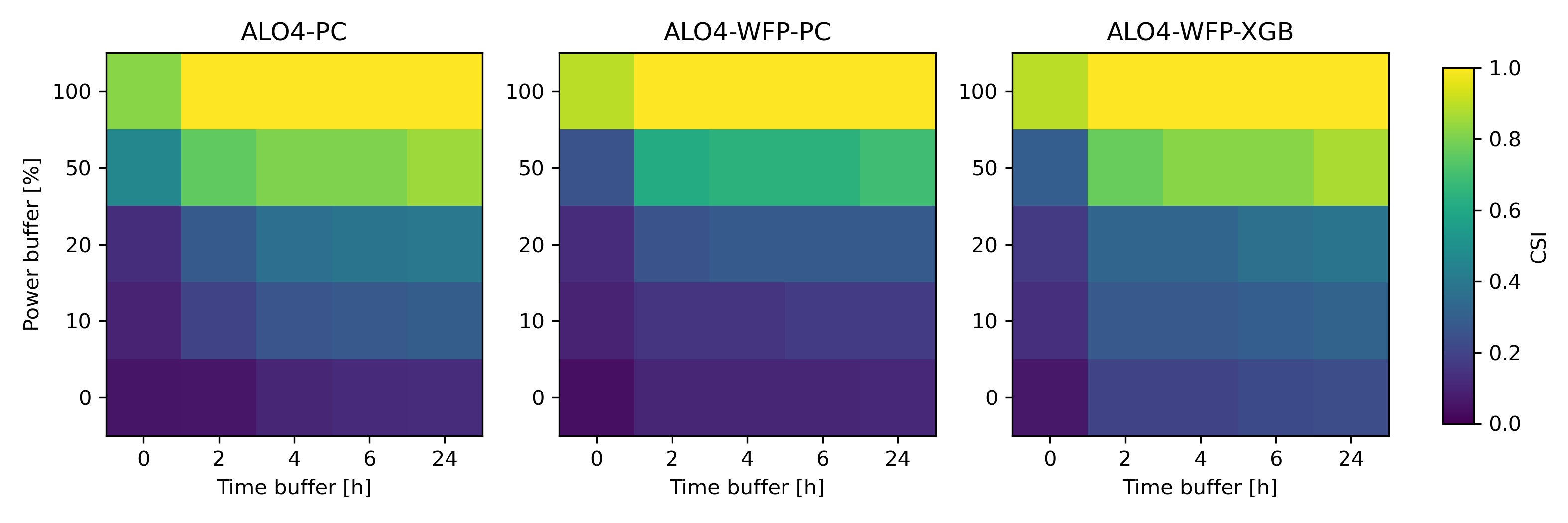} 
    \caption{Intraday CSI for hourly 50\% up-ramping events by different pairs of time and power buffers.}
    \label{fig:CSI_by_buffer_intraday}
\end{figure*}

To illustrate the sensitivity of buffer settings on model evaluation, we take the example of hourly 50\% up-ramping events for discussion (figure \ref{fig:CSI_by_buffer_intraday}). For a fixed time buffer, increasing the power buffer consistently raises CSI scores, and with a 100\% buffer, the CSI approaches 1, though not exactly due to remaining false alarms. Nevertheless, this is a theoretical boundary of the verification framework, and it holds limited operational relevance. As for the time buffer, extending its length consistently increases the CSI, but only up to a certain limit. This indicates that when forecasts substantially underestimate ramping magnitudes, no amount of temporal tolerance can compensate for the error. Taking these buffer sensitivities together, the combined effect of both determines the final CSI score. Ideally, a highly accurate forecasting model should achieve a high CSI score with relatively small power and time buffers. Conversely, if a large buffer is needed to achieve a satisfactory CSI, it suggests a fundamental flaw in the model's predictive performance. Therefore, when evaluating a model, we should not just pursue a high CSI score, but also pay attention to the specific power and time buffers required to achieve that score, as this provides a more complete understanding of the model's strengths and weaknesses.

Although this verification framework can flexibly adjust the buffer, an excessive tolerance for errors solely inflates the hit ratio, but is not operationally favorable. In addition, in situations with frequent, closely successive ramping events, excessively large time buffers risk associating a single predicted event with multiple observed events within the buffer window. This overmatching can lead to a misleading overestimation of hit rates and subsequently inflate forecast skill.

\begin{figure}[htbp]
    \centering
    \begin{subfigure}{0.45\textwidth}
        \centering
        \includegraphics[width=\textwidth]{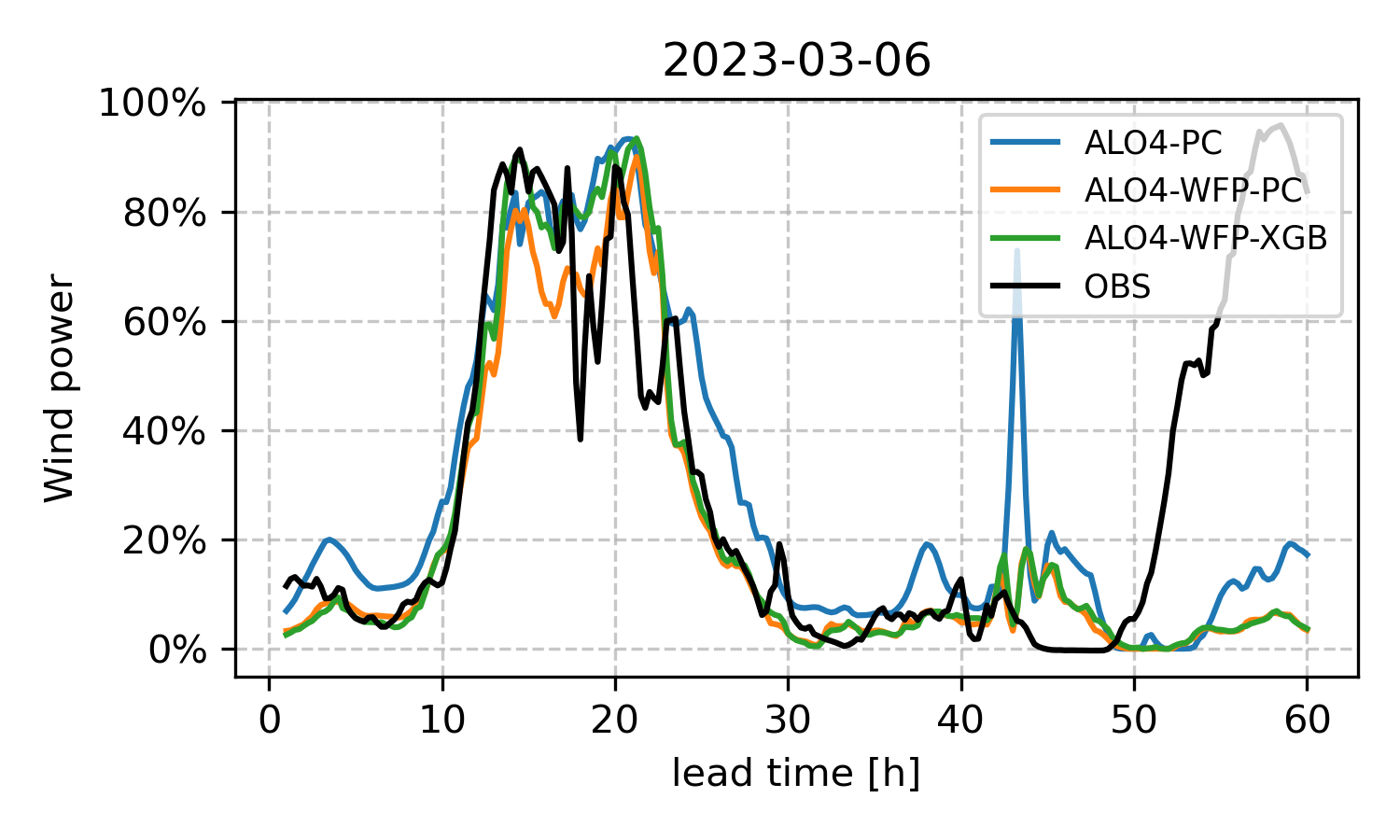} 
    \end{subfigure}
    \hfill
    \begin{subfigure}{0.45\textwidth}
        \centering
        \includegraphics[width=\textwidth]{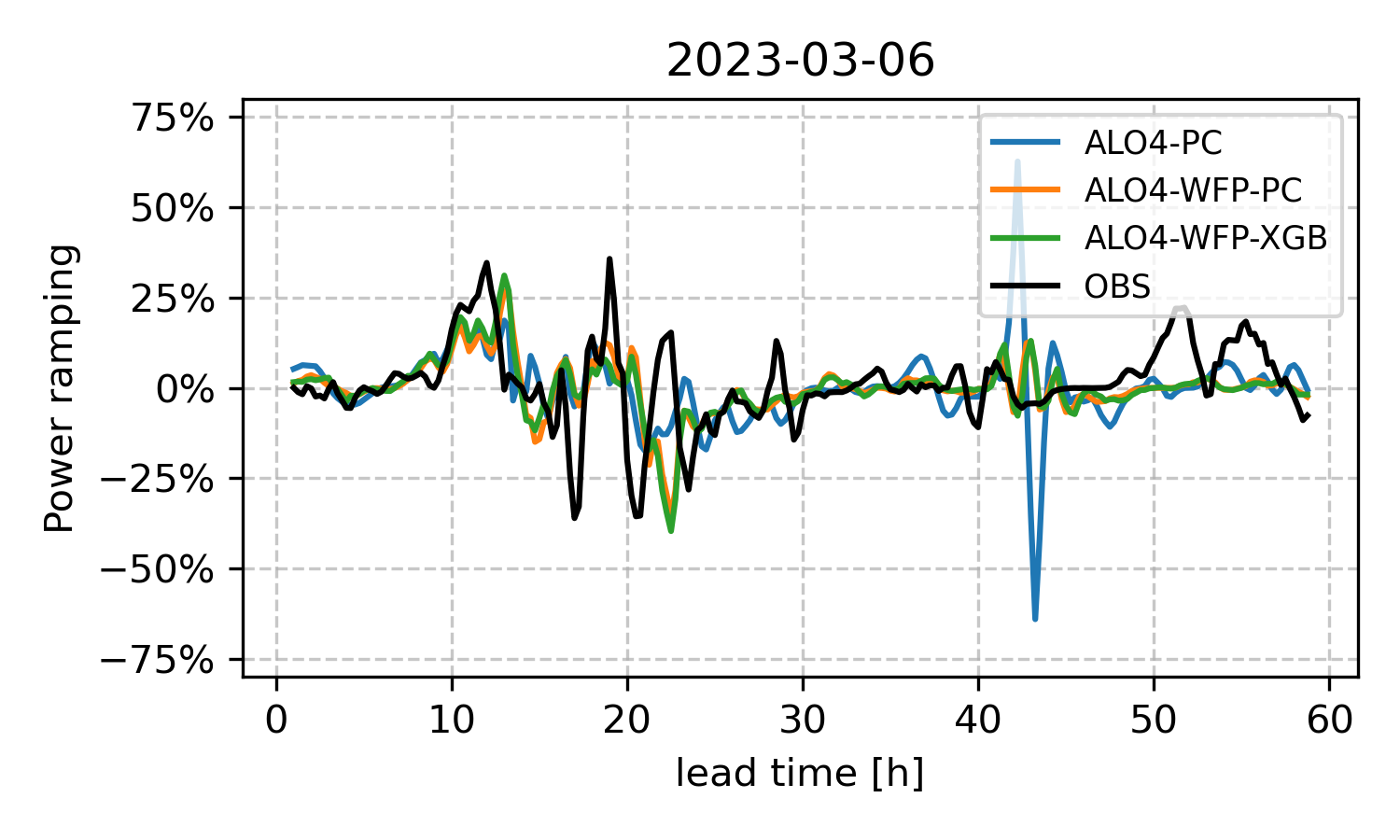} 
    \end{subfigure}
    \caption{Wind power observation and prediction starting from March 6, 2023 (top), and corresponding hourly ramping values (bottom).}
    \label{fig:WP_case}
\end{figure}

This issue of overmatching can be explained by a case study in figure \ref{fig:WP_case}. In this example, the ALO4-WFP-PC model predicted a -35.4\% down-ramp at 22.5 lead hours. At 20.5 lead hours, an observed ramping event of -35.6\% occurred. When verifying a threshold of 30\% ramping with a 2-hour time buffer, the -35.4\% ramping prediction correctly qualifies as a hit, as it falls within the acceptable magnitude range and matches the timing. However, another ramping event of -36.1\% was observed at the 17-hour lead time, and manual inspection confirmed that no model provided a corresponding prediction around that time. As a result, this event was supposed to be classified as a miss. The critical problem would arise if the time buffer is extended to 5.5 hours, where the -36.1\% observed event at 17 hours is retroactively matched with the -35.4\% forecast at 22.5 hours, leading to a false classification of a hit. There are two observed ramping events classified as hits in this case, while only one ramping event was predicted. The ramping event occurring at 17 hours should not be regarded as successfully predicted by the corresponding forecast at 21 hours. In this sense, the predicted ramping event is over-matched with two different observed ramping events. The longer the time buffer, the more this will happen. This example explains the risk of inflating hits when time buffers are overly lenient, particularly under high-frequency ramping conditions. Here we note that under the 1-hour time buffer setting described above, no such overmatching occurs for ramping events exceeding $\pm$15\%. However, a small number of $\pm$10\% ramping events exhibit such cases, as smaller fluctuations occur more frequently. This implies that analyses for small ramping thresholds require more cautious selection of buffer parameters.

\subsection{Event-based verification score for ramping}

The buffer verification framework allows the use of popular metrics in deterministic forecast evaluation, such as the CSI, and provides the flexibility to assess each model's ability in predicting ramping events of varying magnitudes. However, these metrics only count the number of predicted events, without distinguishing the lead time of the event occurrence; whereas the forecast errors are lead-time dependent, as illustrated in the figure \ref{fig:WP_ML_bias_mae}, \ref{fig:WP_ramp_mae}, and table \ref{tab:ramping_MAE}. This motivates our proposal for an event-based scoring rule for the evaluation of ramping event predictability by lead times.

For this purpose, we introduce the Ramp Alignment Score (RAS), a metric designed to quantify the temporal alignment between predicted and observed power ramping events within a specified time window. The score is implemented by splitting the entire lead-time power forecasts into consecutive, non-overlapping time windows of length $T$. Within each window, we define a binary event time index $t$, which captures both the occurrence and the timing of a ramping event. The label of 1 represents the first time step at which the ramp magnitude exceeds a predefined threshold $\theta$, and is assigned a value of 0 if no event occurs within the window. Formally, the observed event time index $t_{obs}$ and the predicted event time index $t_{pred}$ are defined as:

$$
t_{obs} = \begin{cases}
\min\{i \in \{1, ..., T\} \mid |\Delta P_{obs}(i)| \ge \theta\}, & \text{if an event is observed} \\
0, & \text{if no event is observed}
\end{cases}
$$

$$
t_{pred} = \begin{cases}
\min\{i \in \{1, ..., T\} \mid |\Delta P_{pred}(i)| \ge \theta\}, & \text{if an event is predicted} \\
0, & \text{if no event is predicted}
\end{cases}
$$

Using these event time indices, the RAS for each window is calculated as follows:

$$
RAS = \frac{|t_{pred} - t_{obs}|}{T}
$$

This formulation provides an intuitive quantification of the time difference between the observation and prediction. The minimum RAS value of 0 indicates that neither ramping event observations nor predictions occur within the time window, or the predicted ramping event occurs exactly at the same time as the observation, which represents a perfect prediction. A RAS \textgreater 0 indicates (1) the occurrence of either an observed or predicted ramping event within the time window, referring to misses or false alarms; or (2) the presence of both observed and predicted ramping events within the time window but with a temporal displacement, and a larger value means a longer time shift. Therefore, lower RAS values indicate more accurate predictions. 

\begin{figure*}[htbp]
    \centering
    \includegraphics[width=\linewidth]{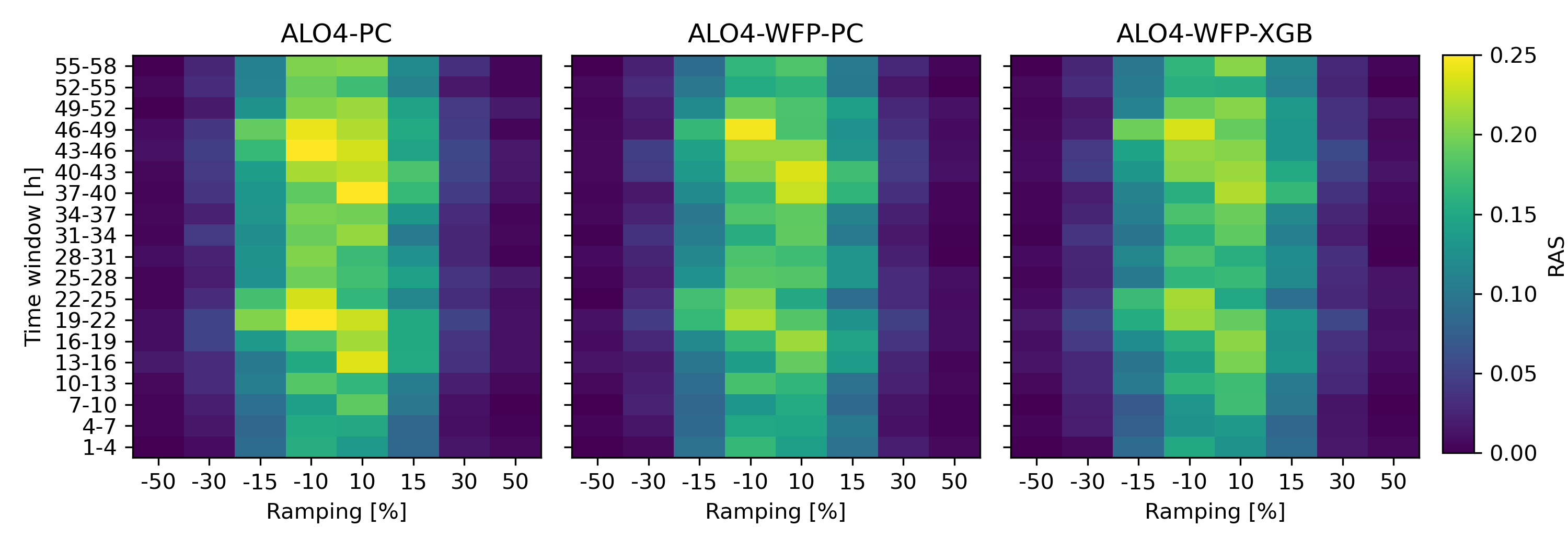} 
    \caption{RAS of BOZ hourly ramping predictions, with a 3-hour-width time window.}
    \label{fig:RAS}
\end{figure*}

We evaluate the RAS of each model across different ramping thresholds with a 3-hour time window (figure \ref{fig:RAS}). This analysis, on the one hand, reveals the temporal displacement between forecasted and observed ramping events within specific time windows, which indicates how ramping predictability differs with lead time. On the other hand, this score can also be used to compare the model's skill. For small ramping events, the predicted and observed events are more likely to coincide within the time window, meaning that prediction errors mainly exist in time shifts rather than failure to detect the event. In contrast, for large ramping events, the average RAS values are close to zero due to their rarity, which limits the opportunity for overlap within any given window. Both ALO4-WFP-PC and ALO4-WFP-XGB models consistently show lower RAS values than ALO4-PC at most thresholds. This suggests that the ALO4-WFP models are more adept at correcting the time shift in ramping events, leading to a higher temporal alignment between forecasts and observations.

\begin{figure}[htbp]
    \centering
    \begin{subfigure}{0.45\textwidth}
        \centering
        \includegraphics[width=\textwidth]{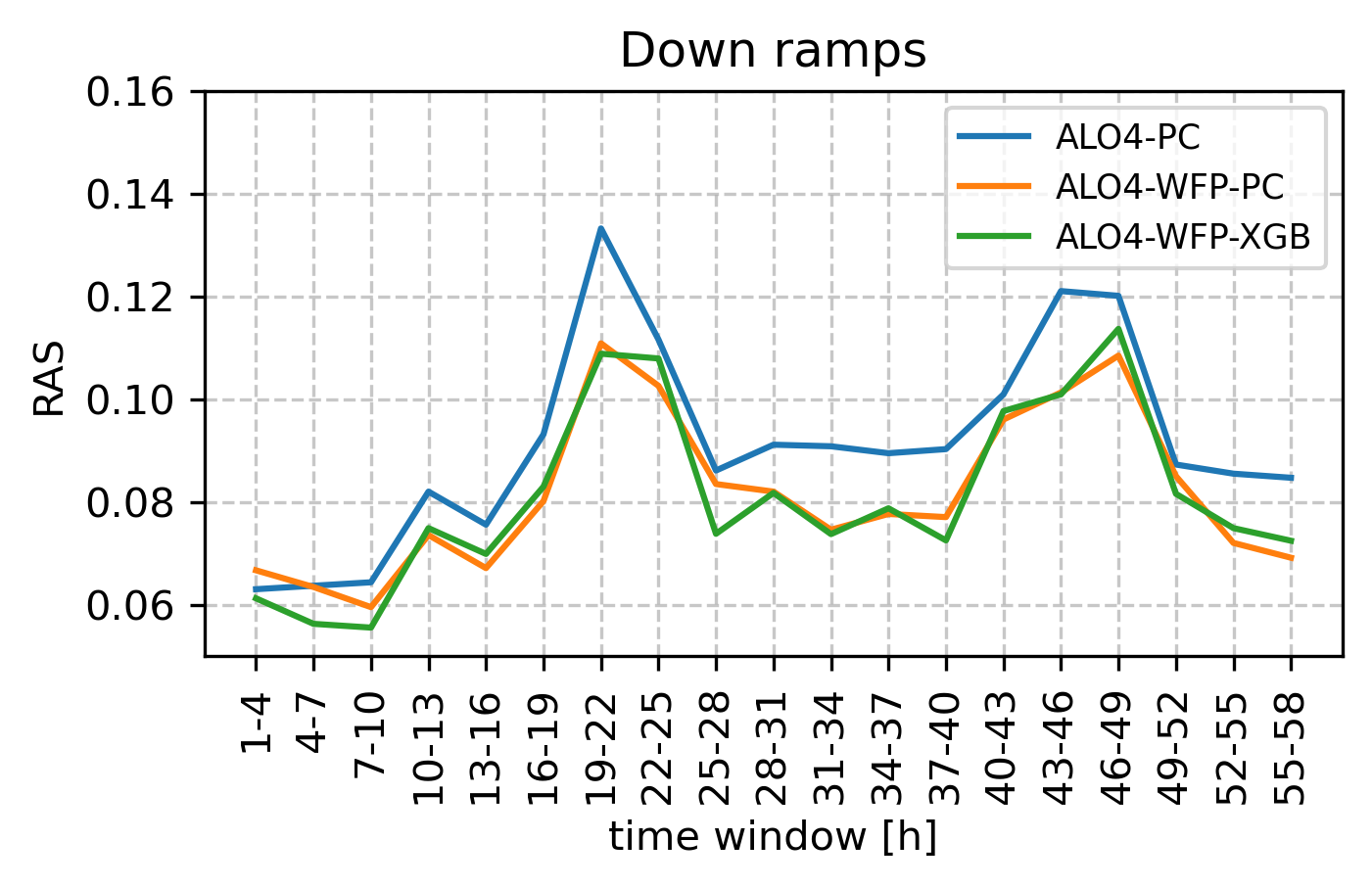} 
    \end{subfigure}
    \hfill
    \begin{subfigure}{0.45\textwidth}
        \centering
        \includegraphics[width=\textwidth]{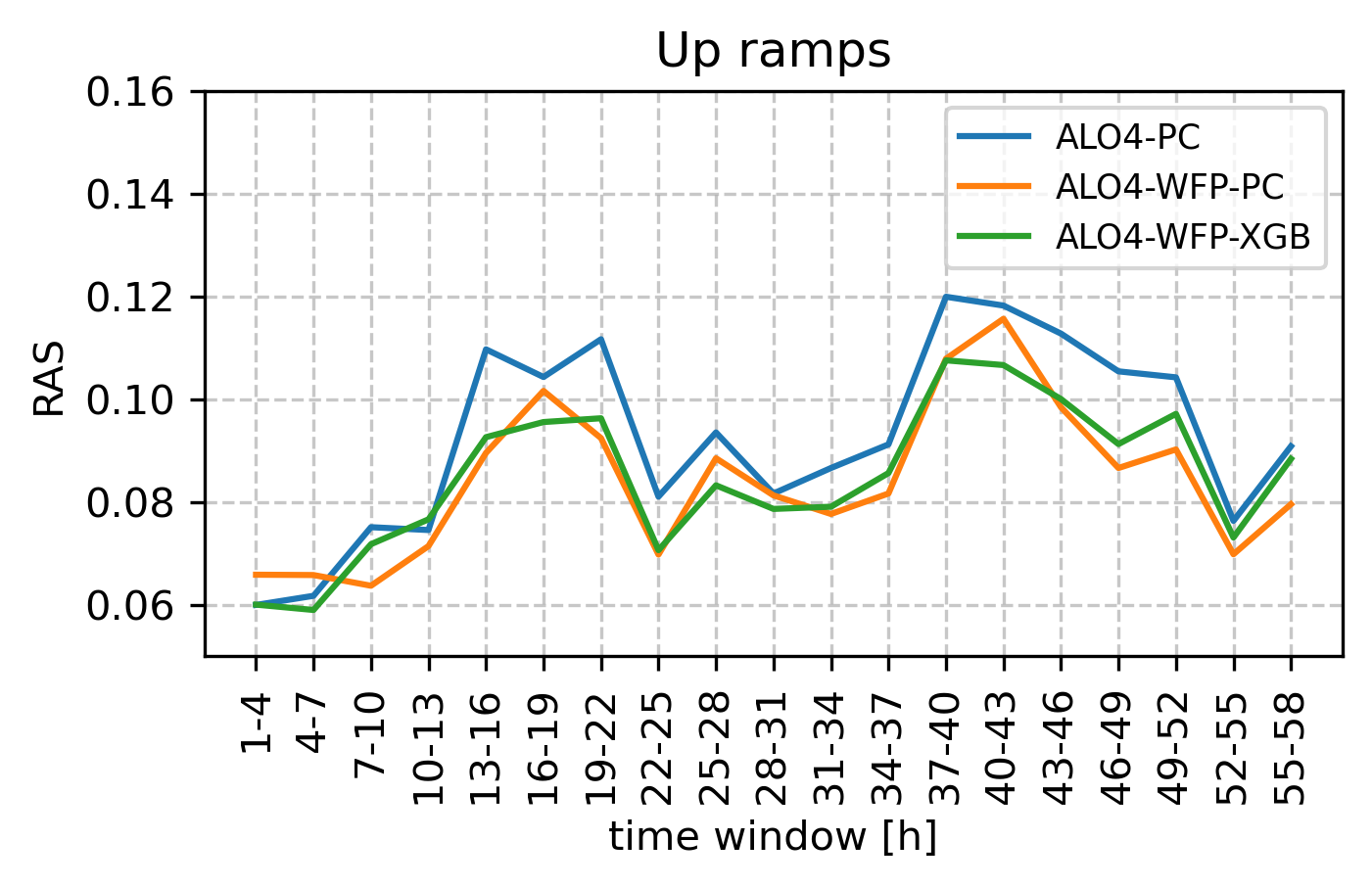} 
    \end{subfigure}
    \caption{RAS averaged overall all of the down (top) and up (bottom) ramping thresholds.}
    \label{fig:RAS_avg}
\end{figure}

The averaged RAS by all thresholds for up and down directions exhibits a pattern of a diurnal cycle (figure \ref{fig:RAS_avg}), resembling those of the ramping MAE (figure \ref{fig:WP_ramp_mae}). The plots show similar RAS of ALO4-WFP-PC and ALO4-WFP-XGB across all lead times, both lower than ALO4-PC, indicating better skill. This reinforces the earlier claim that, although the XGB model achieves a much lower MAE in wind power prediction compared to ALO4-WFP, its improvement in predicting ramping events is not necessarily as significant as in power prediction. The RAS for down-ramping events is also slightly higher than for up-ramping events, suggesting that up-ramping is slightly more predictable. This is consistent with the previous discussion on figure \ref{fig:performance_diagram_intraday}.

\subsection{Impact of precipitation on ramping forecasts}

Understanding how specific meteorological conditions influence the accuracy of ramping forecasts provides more insights into the ramping predictability of models. Among these, precipitation is of particular interest, as it is often associated with convective activity, frontal systems, and other dynamic processes that can induce rapid changes in wind speed and power output \cite{pichault2020characterisation}. In this section, we investigate the conditional verification by 3-hour cumulative precipitation (PCP3h) from ALO4-WFP predictions, aiming to assess whether the level of precipitation systematically relates to model skill or contributes to forecast uncertainty.

\begin{figure}[htbp]
    \centering
    \begin{subfigure}{0.45\textwidth}
        \centering
        \includegraphics[width=\textwidth]{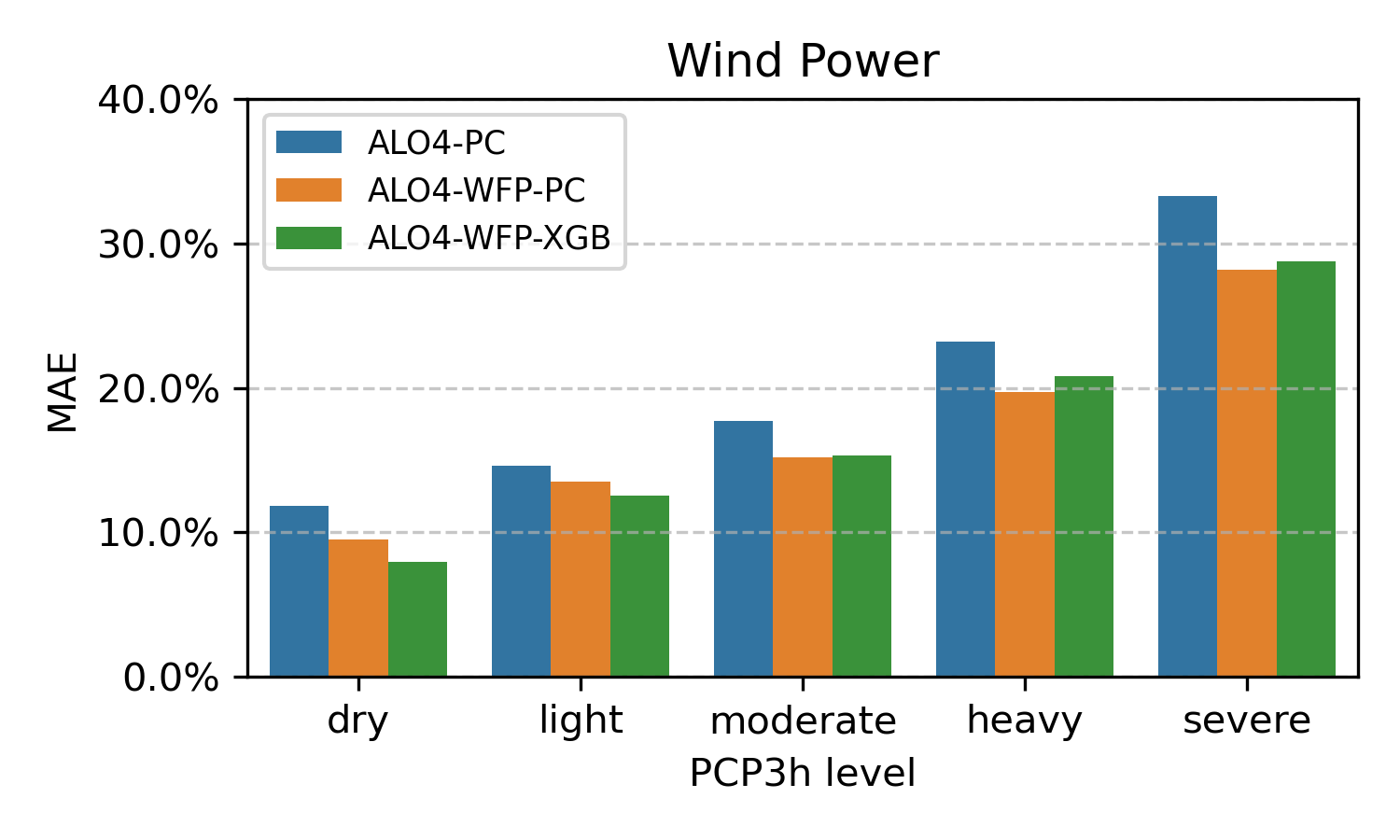} 
    \end{subfigure}
    \hfill
    \begin{subfigure}{0.45\textwidth}
        \centering
        \includegraphics[width=\textwidth]{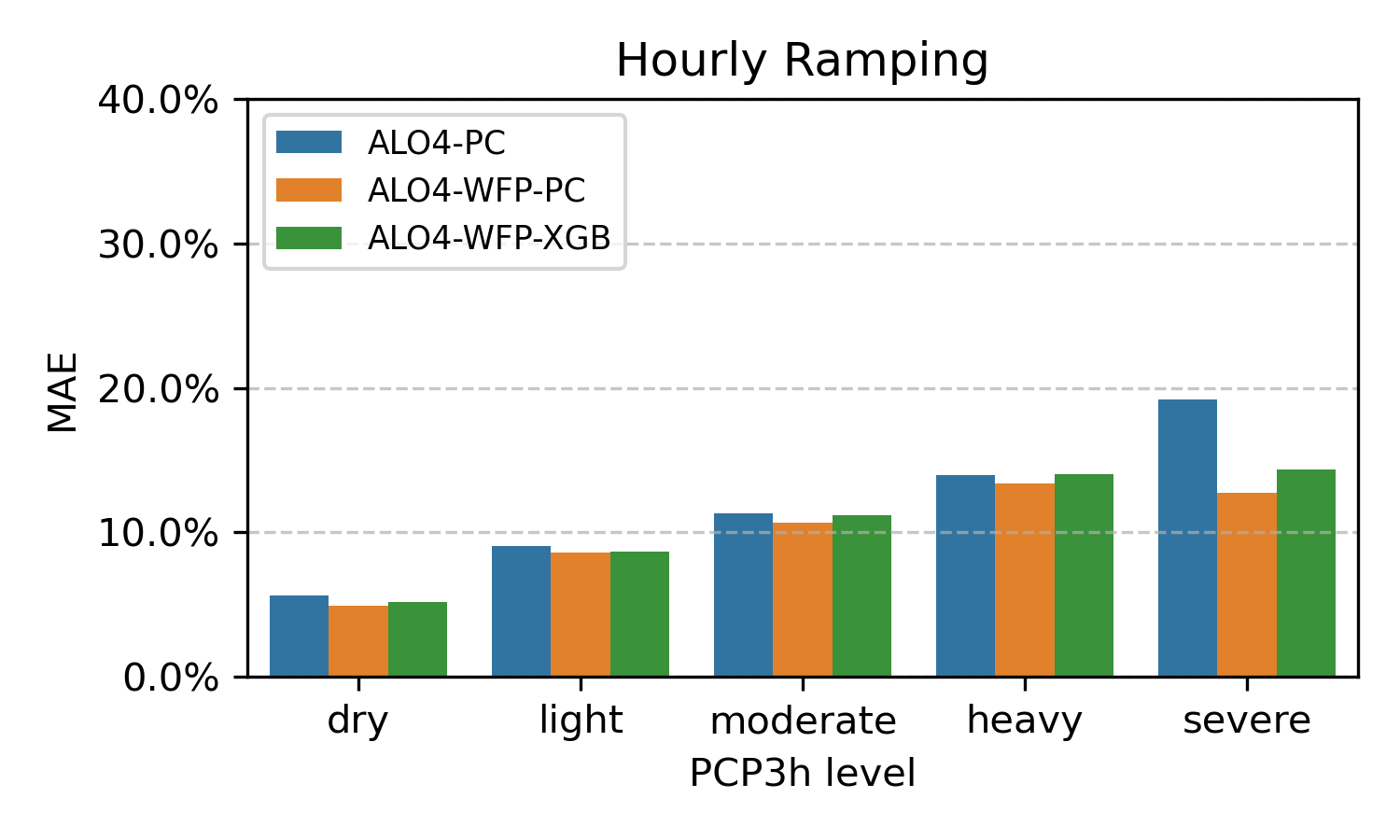} 
    \end{subfigure}
    \caption{MAE of wind power (top) and corresponding hourly ramping (bottom) forecasts by models, expressed as a percentage of BOZ total installed capacity. The results are categorized based on levels of PCP3h from ALO4-WFP forecasts, averaged across all wind farms.}
    \label{fig:PCP_mae}
\end{figure}

The precipitation intensities are classified into five categories based on PCP3h, following the INtegrateD RMI Alert (INDRA) system \cite{Smet2013INDRA}: dry (\textless 0.1 mm), light (0.1-1 mm), moderate (1-5 mm), heavy (5-10 mm), and severe (\textgreater 10 mm). These thresholds correspond approximately to the 60th, 95th, and 99th percentiles of the PCP3h distribution on wet days (PCP3h \textgreater 0.1 mm) in our verification period. We average the power MAE and hourly ramping MAE over all lead times with the condition on precipitation levels (figure \ref{fig:PCP_mae}). The results demonstrate a consistent increase in forecast error with rising precipitation intensity, both in terms of wind power and hourly ramping forecasts.  When comparing the different models, ALO4-WFP forecasts consistently outperform those from the operational ALO4 model across all precipitation levels. Notably, ALO4-WFP-XGB shows the lowest wind power MAE under dry and light precipitation intensity, slightly outperforming ALO4-WFP-PC. However, under moderate, heavy, and severe precipitation, ALO4-WFP-PC performs better, which indicates that the PC may be more reliable than the XGB approach when dealing with higher precipitation scenarios. In terms of hourly ramping prediction, ALO4-WFP-PC shows slightly better performance than ALO4-WFP-XGB across all precipitation levels. This advantage becomes more pronounced under severe precipitation. 

\begin{figure*}[htbp]
    \centering
    \includegraphics[width=\textwidth]{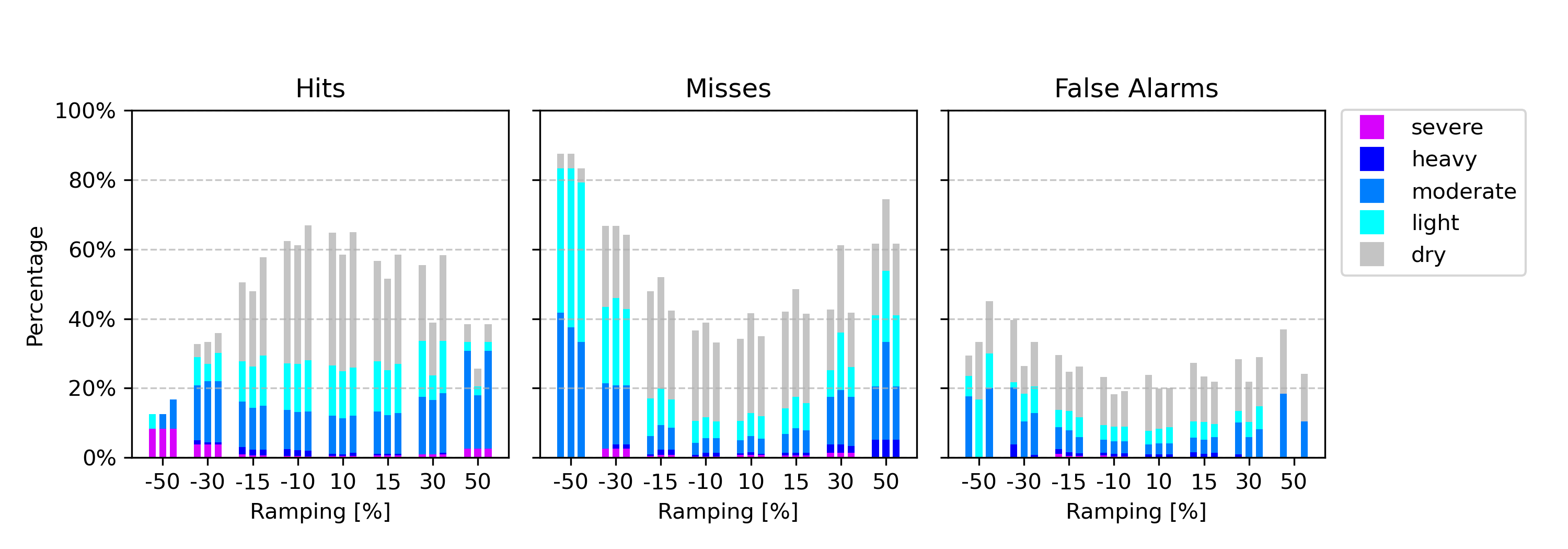}
    \caption{Percentage of intraday ramping event predictions by hits (left), misses (middle), and false alarms (right) to the total number of ramping events by each threshold. Bars are stratified by the PCP3h intensity. The group of bars of each dp respectively represents ALO4-PC, ALO4-WFP-PC, and ALO4-WFP-XGB from left to right. The verification uses 1-hour time buffer and 20\% power buffer.}
    \label{fig:dp_PCP_level_intraday}
\end{figure*}

We categorize the verification outcomes of hits, misses, and false alarms according to the precipitation intensity associated with each event and present the percentage of hourly ramping events for each threshold (figure \ref{fig:dp_PCP_level_intraday}). As in the previous analyses, a 1-hour time buffer and a 20\% power buffer are applied to the verification. The percentages of hits and misses are computed as the number of hits or misses divided by the number of observed ramping events, whereas the percentage of false alarms is calculated as the number of false alarms divided by the number of predicted ramping events. Overall, the results indicate two aspects. First, they confirm that the predictability of large ramping events is generally lower than that of small ones, which is consistent with the conclusions drawn in the previous section. Second, conditioning on PCP3h provides additional insight into how precipitation intensity influences the reliability of model predictions.

Under dry conditions, $\pm$10\% and $\pm$15\% events dominate the distribution. In contrast, in the other ramping thresholds, the share of events associated with precipitation exceeds that of dry cases, suggesting that large ramping events are more frequently linked to precipitation. This association is particularly evident at the -50\% ramping threshold, where the proportion of cases occurring under severe precipitation is significantly higher than for any other ramping magnitude. Notably, nearly all ramping events together with severe precipitation are hits by the predictions, with only a few misses. This indicates that ramping events coinciding with severe precipitation are highly predictable, and such rare but intense precipitation conditions can be regarded as strong signals for the occurrence of big ramping. Although the discussion of figure \ref{fig:PCP_mae} notes a larger MAE under more precipitation conditions, this does not contradict the conclusions here. In addition to the hits, misses, and false alarms shown in the figures \ref{fig:dp_PCP_level_intraday}, there are also correct negatives, i.e., cases where neither observed nor predicted ramping events occur under severe precipitation. In such situations, large errors between forecasts and observations may still be present, but they fall outside the scope of our event-focused discussion. Beyond severe precipitation cases, a considerable number of large ramping events occur under moderate and light precipitation. These events are far less captured by the models, reflecting challenges in forecasting ramping in these intermediate precipitation conditions. Among the false alarms, most $\pm$50\% ramping predictions occur under moderate or light precipitation intensities. This suggests that PCP3h values between 0.1 and 5 mm represent a critical range where model reliability is reduced. We also notice a small number of false-alarmed -15\% ramping events exist under the severe precipitation level. These cases likely correspond to fake ramping signals associated with exceptional precipitation forecasts and would need to be assessed case-by-case.

When comparing the three power models, ALO4-WFP-XGB shows better predictability for $\pm$10\% and $\pm$15\% ramping events under dry conditions. However, this advantage does not extend to precipitation scenarios. Under severe precipitation, all models exhibit similar skill.

\section{Discussion: ramping verification values to operational decision-making}

Better metrics for power prediction do not necessarily reflect improvements in the prediction of ramping events. While the popular measure of skill often revolves around the number of 'hits' for specific magnitude thresholds, achieving precise predictions that perfectly align with predefined, rigid thresholds is exceptionally difficult. In operational practice, a degree of tolerance is inherent; a forecast that predicts a ramp slightly lagging the observation or slightly underestimating its magnitude is not a complete failure and can still provide significant value for grid management. This discrepancy between rigid verification and practical utility motivated our development of an appropriate verification framework for ramping events. Such a framework aims to be sufficiently flexible to incorporate operational tolerances, such as user-defined buffers for time and power; furthermore, to be extensible, allowing users to tailor the verification process to their specific needs and risk profiles.

The buffer verification framework provides an intuitive evaluation of ramping forecasts by incorporating time and power deviation tolerances. The buffer approach reflects how system operators respond in practice. For example, treating a ramp predicted 15 minutes early as still actionable when sufficient energy resources are available. This enables a more flexible assessment of forecast utility, particularly in contexts where timely and reliable ramp identification is critical. Importantly, buffer verification demonstrates the predictability of ramping events of varying intensities and effectively reveals the strengths and weaknesses of each model in practical applications. Our findings show that small, frequent ramps are more predictable, while large and extreme ramping events remain challenging, with low CSI even with buffers. This indicates that the prediction of extreme ramping remains a challenge in power dispatch.

The accuracy of ramping forecasts carries significant implications. Accurate predictions reduce balancing costs, improve market efficiency, and enhance system reliability, whereas missed events lead to costly emergency measures and threaten grid stability. False alarms, though less damaging, still impose avoidable costs and erode operator confidence \cite{hasan2025state}. Decision-making is supposed to weigh statistical skill against operational priorities, balancing risk, reliability, and cost of misses and false alarms. Our study reveals that the low power variability of ALO-WFP-PC can reduce false alarms, but it also leads to more misses. Our study helps to understand the predictive performance, thereby providing a practical reference for model selection.

The proposed event-based RAS is designed to evaluate the timing shifts that commonly occur between predicted and observed ramping events. The time window explicitly reflects the tolerance of timing errors that can be operationally acceptable. Compared with MAE, the strength of RAS is the ability to verify events at specific ramping intensities and to benchmark performance across models. As discussed earlier, model-derived ramping MAE alone cannot distinguish models' skill, whereas RAS clearly demonstrates that power models based on ALO4-WFP outperform those based on ALO4 across all lead times. Nevertheless, the current definition of RAS does not yet account for magnitude errors. This represents a promising direction for future development to further enhance its operational relevance.

\section{Conclusions and outlook}

The study investigates wind power and ramping event prediction skill for the Belgian Offshore Zone. We first introduce NWP models, including RMI's operational ALO4, the wind farm parameterization enhanced model ALO4-WFP, and ECMWF's HRES. Wind power predictions are obtained by converting the NWP models’ wind speed forecasts with the power curve. The results show that ALO4-WFP-PC achieved the lowest MAE across all lead times, and a negative bias is found for ALO4-WFP-PC. We also present HRES as a model reference. Although its MAE at lead times of more than 20 hours is lower than that of ALO4-PC, it remains higher than ALO4-WFP-PC, underscoring the added value of the WFP in offshore forecasting. Considering the limitations of simple power curve conversion, we apply machine learning methods, specifically neural networks and XGBoost, for power prediction. These approaches successfully correct bias to near zero and yielded lower MAE values than both ALO4-PC and ALO4-WFP-PC.

However, better average power prediction scores do not inherently guarantee higher skill in forecasting power ramps. Although ML models lead to lower power MAE, their ramping values corresponding to these power predictions have very similar MAE to those of ALO4-WFP-PC, and the average of all lead times is even slightly higher than that of ALO4-WFP-PC. The PSD analysis indicates that ALO4-PC tends to overestimate ramping intensity on timescales longer than one hour, whereas ALO4-WFP-PC exhibits insufficient variability in its power predictions. ML models produce variability that more closely matches observations. However, in hourly and shorter timescales, all models underestimate the ramping intensity. These underestimations highlight the challenge of accurately forecasting short-term ramping events, which are of particular operational concern.

Forecasts may fail to capture the timing or magnitude of rapid power fluctuations, thereby undermining their value in real-time dispatch or reserve scheduling. This motivates the need for evaluation methods that go beyond statistical error analysis and directly assess the decision-making relevance of forecasts, especially for ramping events. We develop a flexible verification framework based on the traditional contingency table to evaluate ramping event predictability by incorporating parameters, including magnitude, duration, and time and power buffers to account for prediction error tolerances. This approach enables a more realistic evaluation of forecast utility by allowing for minor deviations in timing or magnitude. Our verification compares the forecasting skill of different models across multiple ramping thresholds under two configurations: strict verification criteria and loose verification with time and power buffers. On the one hand, applying buffers provides a more flexible evaluation framework, which allows the predictability of ramping events to be revealed more clearly. On the other hand, it improves a fair comparison between models by reducing the impact of slight timing or magnitude mismatches. The results show that, compared with ALO4-PC, ALO4-WFP-PC substantially reduces the number of false alarms, at the cost of a higher number of misses. ALO4-WFP-XGB, while still having errors, achieves a slightly higher CSI than both ALO4-PC and ALO4-WFP-PC. A consistent finding across all models was the higher predictability of up-ramping events compared to down-ramping events.

We additionally examine the impact of buffer settings on verification scores. While more time and power buffers increase hit rates by allowing more tolerance, too lenient buffers lead to over-inflated scores and falsely match unrelated events, reducing the reference value of the results. 

Furthermore, we propose an event-based scoring rule, the Ramp Alignment Score (RAS), which quantifies the alignment between predicted and observed power ramping events within a specified time window. The advantage of RAS analysis is that it complements the defect that contingency table validation ignores the lead time dimension. We apply a 3-hour time window and calculate the RAS for all models over all lead times. The result reveals a diurnal cycle in ramping predictability and confirms that both ALO4-WFP-PC and ALO4-WFP-XGB outperform ALO4-PC in predicting ramping events.

We also investigate the indicator of precipitation forecasts on wind power ramp predictability. The analysis suggests that under the scenarios of greater precipitation intensity, the prediction errors of each model become larger. ALO4-WFP-XGB provides more accurate forecasts under dry and light precipitation conditions, whereas ALO4-WFP-PC has the lowest MAE in moderate-to-severe precipitation scenarios. The ramping verification demonstrates that large ramping events are associated with precipitation, especially under severe conditions where model skill is notably higher and most events are successfully captured. Conversely, ramping events accompanied by moderate and light precipitation have a high percentage of misses and false alarms, especially in big ramping events. This finding illustrates that extreme rainfall provides strong meteorological signals for large, highly predictable ramps, while scenarios of intermediate precipitation levels remain a challenge for successful ramp forecasting.

In conclusion, this research provides a comprehensive assessment of wind power ramping predictability. The developed verification framework with buffers enables flexibility in error tolerance and benchmarking. The proposed RAS serves as an additional metric that provides insight into the models' performance by lead time. Optimizing power prediction under different precipitation intensity scenarios is a possible direction to improve the predictability of ramping events.

In future work, we expect to incorporate precipitation (and even more meteorological variables) into the model inputs for power prediction, as this study demonstrates that ramping event prediction varies across precipitation intensities. In the current setup, ramping forecasts are derived from analyses of independently predicted time series at each lead time. In the following, we plan to employ Transformer-based architectures to process the entire lead-time sequence jointly \cite{van2025self}, aiming to improve both power and ramping predictability with the proposed verification framework. Besides, a more comprehensive conditional verification will investigate the ramping event predictions under various meteorological conditions and their associated synoptic-scale dynamics, thereby contributing to a deeper understanding of power ramping. Additionally, the present study focuses exclusively on deterministic forecasts, while probabilistic forecasts offer a valuable complementary perspective by quantifying uncertainty for ramp events. This motivates the development of probabilistic ramp forecasting, for instance, through the statistical post-processing of ensembles \cite{muschinski2022predicting}. A key challenge is effectively modeling the temporal dependencies across lead times, which are crucial for ramp events. This shift toward probabilistic forecasting also requires appropriate verification strategies tailored to ramping events, such as the Brier Skill Score (BSS), to properly assess forecast skill.

\section{Acknowledgments}
The authors extend their appreciation to the Belgian Science Policy Office (BELSPO) for providing financial support for this project (B2/223/P1/E-TREND). The authors also acknowledge the BeFORECAST project, which is supported by the Energy Transition Fund of the Belgian Federal Government, for the availability of the ALARO forecasts with Fitch wind farm parameterization and power forecasts with neural network used in this paper.

\section{Funding}
This work was supported by the Belgian Science Policy Office (BELSPO) (B2/223/P1/E-TREND).

\clearpage

\bibliography{reference}

\begin{thebibliography}{10}

\bibitem{murcia2020power}
Juan~Pablo Murcia~Leon, Matti~Juhani Koivisto, Poul S{\o}rensen, and Philippe Magnant.
\newblock Power fluctuations in high installation density offshore wind fleets.
\newblock {\em Wind Energy Science Discussions}, 2020:1--23, 2020.

\bibitem{elia2018offshore}
Elia.
\newblock Analysis, benchmark and mitigation of storm and ramping risks from offshore wind power in belgium.
\newblock Technical report, Elia, 2018.

\bibitem{elia2026adequacy}
Elia.
\newblock Adequacy and flexibility study for belgium (2026-2036).
\newblock Technical report, Elia, 2025.

\bibitem{kariniotakis2017renewable}
Georges Kariniotakis.
\newblock {\em Renewable energy forecasting: from models to applications}.
\newblock Woodhead Publishing, 2017.

\bibitem{gonzalez2012wake}
Francisco Gonz{\'a}lez-Longatt, Peter Wall, and Vladimir Terzija.
\newblock Wake effect in wind farm performance: Steady-state and dynamic behavior.
\newblock {\em Renewable Energy}, 39(1):329--338, 2012.

\bibitem{archer2018review}
Cristina~L Archer, Ahmadreza Vasel-Be-Hagh, Chi Yan, Sicheng Wu, Yang Pan, Joseph~F Brodie, and A~Eoghan Maguire.
\newblock Review and evaluation of wake loss models for wind energy applications.
\newblock {\em Applied Energy}, 226:1187--1207, 2018.

\bibitem{fischereit2022review}
Jana Fischereit, Roy Brown, Xiaoli~Guo Lars{\'e}n, Jake Badger, and Graham Hawkes.
\newblock Review of mesoscale wind-farm parametrizations and their applications.
\newblock {\em Boundary-Layer Meteorology}, 182(2):175--224, 2022.

\bibitem{van2022one}
Bart van Stratum, Natalie Theeuwes, Jan Barkmeijer, Bert van Ulft, and Ine Wijnant.
\newblock A one-year-long evaluation of a wind-farm parameterization in harmonie-arome.
\newblock {\em Journal of Advances in Modeling Earth Systems}, 14(7):e2021MS002947, 2022.

\bibitem{fitch2012local}
Anna~C Fitch, Joseph~B Olson, Julie~K Lundquist, Jimy Dudhia, Alok~K Gupta, John Michalakes, and Idar Barstad.
\newblock Local and mesoscale impacts of wind farms as parameterized in a mesoscale nwp model.
\newblock {\em Monthly Weather Review}, 140(9):3017--3038, 2012.

\bibitem{dieter2025improving}
Dieter Van~den Bleeken, Geert Smet, Joris Van~den Bergh, Idir Dehmous, Daan Degrauwe, Michiel Van~Ginderachter, and Alex Deckmyn.
\newblock Improving wind power forecasts in the {B}elgian {N}orth {S}ea with a wind farm parameterization and a neural network.
\newblock {\em Accepted for publication in Advances in Science and Research}, 2025.

\bibitem{greaves2009temporal}
Beatrice Greaves, Jonathan Collins, Jeremy Parkes, and Andrew Tindal.
\newblock Temporal forecast uncertainty for ramp events.
\newblock {\em Wind Engineering}, 33(4):309--319, 2009.

\bibitem{drew2018identifying}
Daniel~R Drew, Janet~F Barlow, and Phil~J Coker.
\newblock Identifying and characterising large ramps in power output of offshore wind farms.
\newblock {\em Renewable Energy}, 127:195--203, 2018.

\bibitem{pichault2020characterisation}
Mathieu Pichault, Claire Vincent, Grant Skidmore, and Jason Monty.
\newblock Characterisation of intra-hourly wind power ramps at the wind farm scale and associated processes.
\newblock {\em Wind Energy Science Discussions}, 2020:1--25, 2020.

\bibitem{gallego2015review}
Cristobal Gallego-Castillo, Alvaro Cuerva-Tejero, and Oscar Lopez-Garcia.
\newblock A review on the recent history of wind power ramp forecasting.
\newblock {\em Renewable and Sustainable Energy Reviews}, 52:1148--1157, 2015.

\bibitem{potter2009potential}
Cameron~W Potter, Eric Grimit, and Bart Nijssen.
\newblock Potential benefits of a dedicated probabilistic rapid ramp event forecast tool.
\newblock In {\em 2009 IEEE/PES Power Systems Conference and Exposition}, pages 1--5. IEEE, 2009.

\bibitem{cutler2007detecting}
Nicholas Cutler, Merlinde Kay, Kieran Jacka, and Torben~Skov Nielsen.
\newblock Detecting, categorizing and forecasting large ramps in wind farm power output using meteorological observations and wppt.
\newblock {\em Wind Energy}, 10(5):453--470, 2007.

\bibitem{jin2024evaluation}
Chenxi Jin, Yang Yang, Chao Han, Ting Lei, Chen Li, and Bing Lu.
\newblock Evaluation of forecasted wind speed at turbine hub height and wind ramps by five nwp models with observations from 262 wind farms over china.
\newblock {\em Meteorological Applications}, 31(6):e70007, 2024.

\bibitem{vallance2017towards}
Lo{\"\i}c Vallance, Bruno Charbonnier, Nicolas Paul, St{\'e}phanie Dubost, and Philippe Blanc.
\newblock Towards a standardized procedure to assess solar forecast accuracy: A new ramp and time alignment metric.
\newblock {\em Solar Energy}, 150:408--422, 2017.

\bibitem{messner2020evaluation}
Jakob~W Messner, Pierre Pinson, Jethro Browell, Mathias~B Bjerreg{\aa}rd, and Irene Schicker.
\newblock Evaluation of wind power forecasts—an up-to-date view.
\newblock {\em Wind Energy}, 23(6):1461--1481, 2020.

\bibitem{bianco2016wind}
Laura Bianco, Irina~V Djalalova, James~M Wilczak, Joel Cline, Stan Calvert, Elena Konopleva-Akish, Cathy Finley, and Jeffrey Freedman.
\newblock A wind energy ramp tool and metric for measuring the skill of numerical weather prediction models.
\newblock {\em Weather and Forecasting}, 31(4):1137--1156, 2016.

\bibitem{Kamath5484508}
Chandrika Kamath.
\newblock Understanding wind ramp events through analysis of historical data.
\newblock In {\em IEEE PES T\&D 2010}, pages 1--6, 2010.

\bibitem{bradford2010forecasting}
Kristen~T Bradford, RL~Carpenter, and Brent Shaw.
\newblock Forecasting southern plains wind ramp events using the wrf model at 3-km.
\newblock In {\em AMS Student Conference}, volume~1, pages 99--114, 2010.

\bibitem{zhang2017ramp}
Jie Zhang, Mingjian Cui, Bri-Mathias Hodge, Anthony Florita, and Jeffrey Freedman.
\newblock Ramp forecasting performance from improved short-term wind power forecasting over multiple spatial and temporal scales.
\newblock {\em Energy}, 122:528--541, 2017.

\bibitem{cui2023algorithm}
Yang Cui, Zhenghong Chen, Yingjie He, Xiong Xiong, and Fen Li.
\newblock An algorithm for forecasting day-ahead wind power via novel long short-term memory and wind power ramp events.
\newblock {\em Energy}, 263:125888, 2023.

\bibitem{gallego2013wavelet}
Crist{\'o}bal~Jos{\'e} Gallego~Castillo, Alexandre Costa, Alvaro Cuerva~Tejero, Lars Landberg, Beatrice Greaves, and Jonathan Collins.
\newblock A wavelet-based approach for large wind power ramp characterisation.
\newblock {\em Wind Energy}, 16(2):257--278, 2013.

\bibitem{cheneka2020simple}
Bedassa~R Cheneka, Simon~J Watson, and Sukanta Basu.
\newblock A simple methodology to detect and quantify wind power ramps.
\newblock {\em Wind Energy Science}, 5(4):1731--1741, 2020.

\bibitem{bossavy2015edge}
Arthur Bossavy, Robin Girard, and George Kariniotakis.
\newblock An edge model for the evaluation of wind power ramps characterization approaches.
\newblock {\em Wind Energy}, 18(7):1169--1184, 2015.

\bibitem{zhang2022short}
Wanqing Zhang, Zi~Lin, and Xiaolei Liu.
\newblock Short-term offshore wind power forecasting-a hybrid model based on discrete wavelet transform (dwt), seasonal autoregressive integrated moving average (sarima), and deep-learning-based long short-term memory (lstm).
\newblock {\em Renewable Energy}, 185:611--628, 2022.

\bibitem{frias2016introducing}
Laura Fr{\'\i}as-Paredes, Ferm{\'\i}n Mallor, Teresa Le{\'o}n, and Mart{\'\i}n Gast{\'o}n-Romeo.
\newblock Introducing the temporal distortion index to perform a bidimensional analysis of renewable energy forecast.
\newblock {\em Energy}, 94:180--194, 2016.

\bibitem{termonia2018aladin}
Piet Termonia, Claude Fischer, Eric Bazile, Fran{\c{c}}ois Bouyssel, Radmila Bro{\v{z}}kov{\'a}, Pierre B{\'e}nard, Bogdan Bochenek, Daan Degrauwe, Mari{\'a} Derkov{\'a}, Ryad El~Khatib, et~al.
\newblock The aladin system and its canonical model configurations arome cy41t1 and alaro cy40t1.
\newblock {\em Geoscientific Model Development}, 11(1):257--281, 2018.

\bibitem{carrillo2013review}
C~Carrillo, AF~Obando Monta{\~n}o, J~Cidr{\'a}s, and Eloy D{\'\i}az-Dorado.
\newblock Review of power curve modelling for wind turbines.
\newblock {\em Renewable and Sustainable Energy Reviews}, 21:572--581, 2013.

\bibitem{chen2016xgboost}
Tianqi Chen and Carlos Guestrin.
\newblock Xgboost: A scalable tree boosting system.
\newblock {\em Proceedings of the 22nd ACM SIGKDD International Conference on Knowledge Discovery and Data Mining}, pages 785--794, 2016.

\bibitem{ferreira2011survey}
Carlos Ferreira, Joao Gama, L~Matias, Audun Botterud, and J\_ Wang.
\newblock A survey on wind power ramp forecasting.
\newblock Technical report, Argonne National Lab.(ANL), Argonne, IL (United States), 2011.

\bibitem{lee2012analyzing}
Duehee Lee and Ross Baldick.
\newblock Analyzing the variability of wind power output through the power spectral density.
\newblock In {\em 2012 IEEE Power and Energy Society General Meeting}, pages 1--8, 2012.

\bibitem{larsen2012recipes}
Xiaoli~Guo Lars{\'e}n, S{\o}ren Ott, Jake Badger, Andrea~N Hahmann, and Jakob Mann.
\newblock Recipes for correcting the impact of effective mesoscale resolution on the estimation of extreme winds.
\newblock {\em Journal of Applied Meteorology and Climatology}, 51(3):521--533, 2012.

\bibitem{roebber2009visualizing}
Paul~J Roebber.
\newblock Visualizing multiple measures of forecast quality.
\newblock {\em Weather and Forecasting}, 24(2):601--608, 2009.

\bibitem{zack2006optimization}
J~W Zack.
\newblock Optimization of wind power production forecast performance during critical periods for grid management, 2006.

\bibitem{Smet2013INDRA}
Geert Smet, Joris Van~den Bergh, Maarten Reyniers, Dieter Poelman, Daan Degrauwe, Alex Deckmyn, Rafiq Hamdi, Piet Termonia, and Laurent Delobbe.
\newblock The integrated rmi alert system (indra).
\newblock In {\em EMS Annual Meeting Abstracts}, volume~10, 2013.

\bibitem{hasan2025state}
Mahmudul Hasan, Zannatul Mifta, Sumaiya~Janefar Papiya, Paromita Roy, Pronay Dey, Nafisa~Atia Salsabil, Nahid-Ur-Rahman Chowdhury, and Omar Farrok.
\newblock A state-of-the-art comparative review of load forecasting methods: Characteristics, perspectives, and applications.
\newblock {\em Energy Conversion and Management: X}, 26:100922, 2025.

\bibitem{van2025self}
Aaron Van~Poecke, Tobias~Sebastian Finn, Ruoke Meng, Joris Van~den Bergh, Geert Smet, Jonathan Demaeyer, Piet Termonia, Hossein Tabari, and Peter Hellinckx.
\newblock Self-attentive transformer for fast and accurate postprocessing of temperature and wind speed forecasts.
\newblock {\em Artificial Intelligence for the Earth Systems}, 4(3):240127, 2025.

\bibitem{muschinski2022predicting}
Thomas Muschinski, Moritz~N Lang, Georg~J Mayr, Jakob~W Messner, Achim Zeileis, and Thorsten Simon.
\newblock Predicting power ramps from joint distributions of future wind speeds.
\newblock {\em Wind Energy Science}, 7(6):2393--2405, 2022.

\end{thebibliography}

\clearpage

\appendix 

\section*{Appendix: supplementary figures for day-ahead results}

\noindent Note: Supplementary figures are labeled as figure 1A, 2A, etc., to correspond to figures 1, 2, etc., in the main text.

\begin{figure}[ht]
    \centering
    \begin{subfigure}{\linewidth} 
        \centering
        \includegraphics[width=\linewidth]{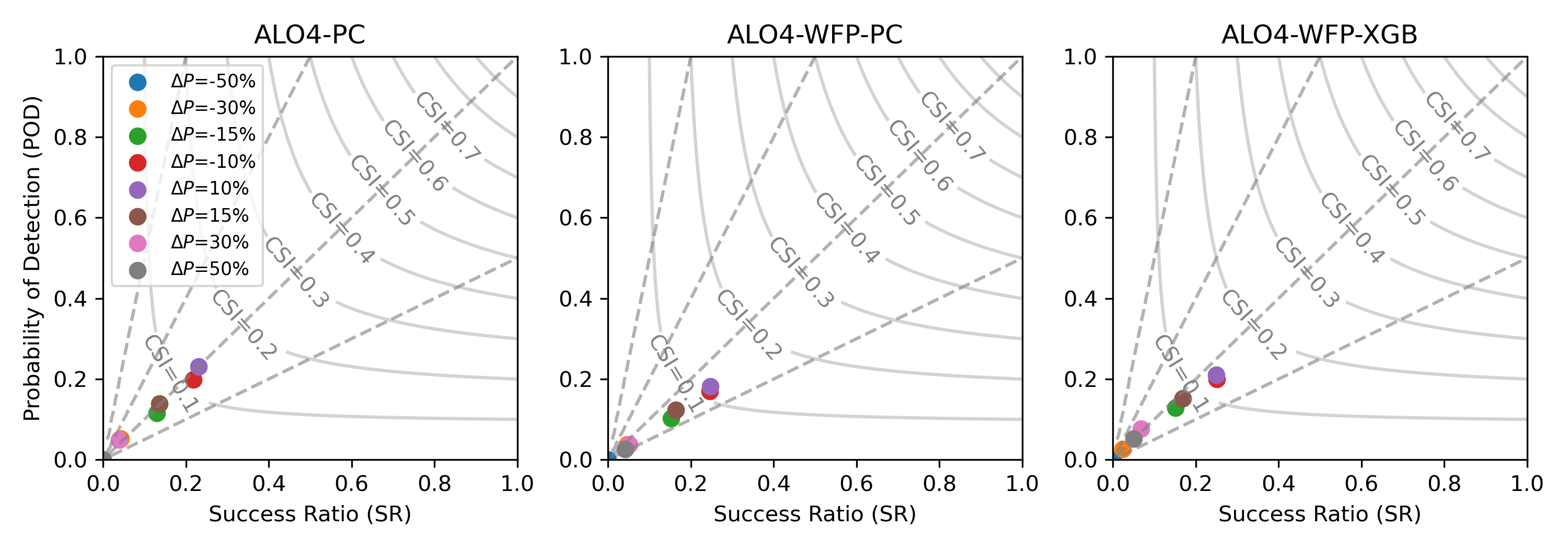} 
    \end{subfigure}

    \begin{subfigure}{\linewidth} 
        \centering
        \includegraphics[width=\linewidth]{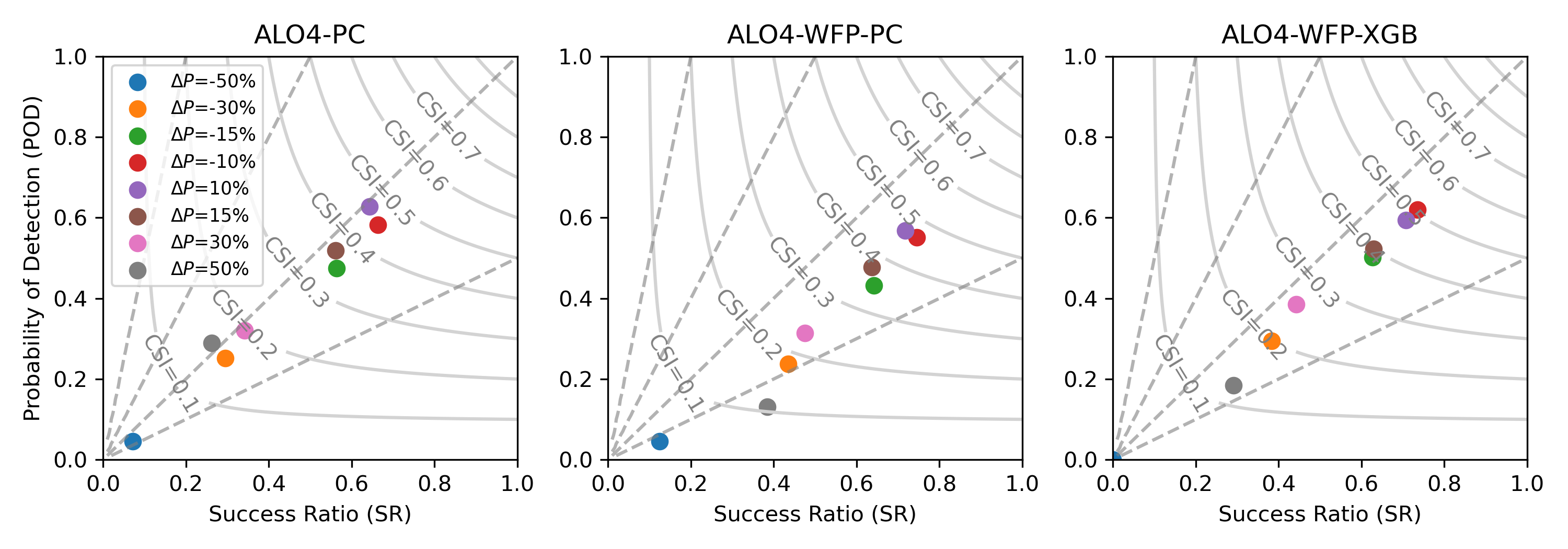} 
    \end{subfigure}

    \caption*{Figure \ref{fig:performance_diagram_intraday}A: Performance diagrams of BOZ ramping event day-ahead predictions. The buffers are the same as those in figure \ref{fig:performance_diagram_intraday}.}
    \label{fig:performance_diagram_dayahead}
\end{figure}

\begin{figure}[ht]
    \centering
    \includegraphics[width=\linewidth]{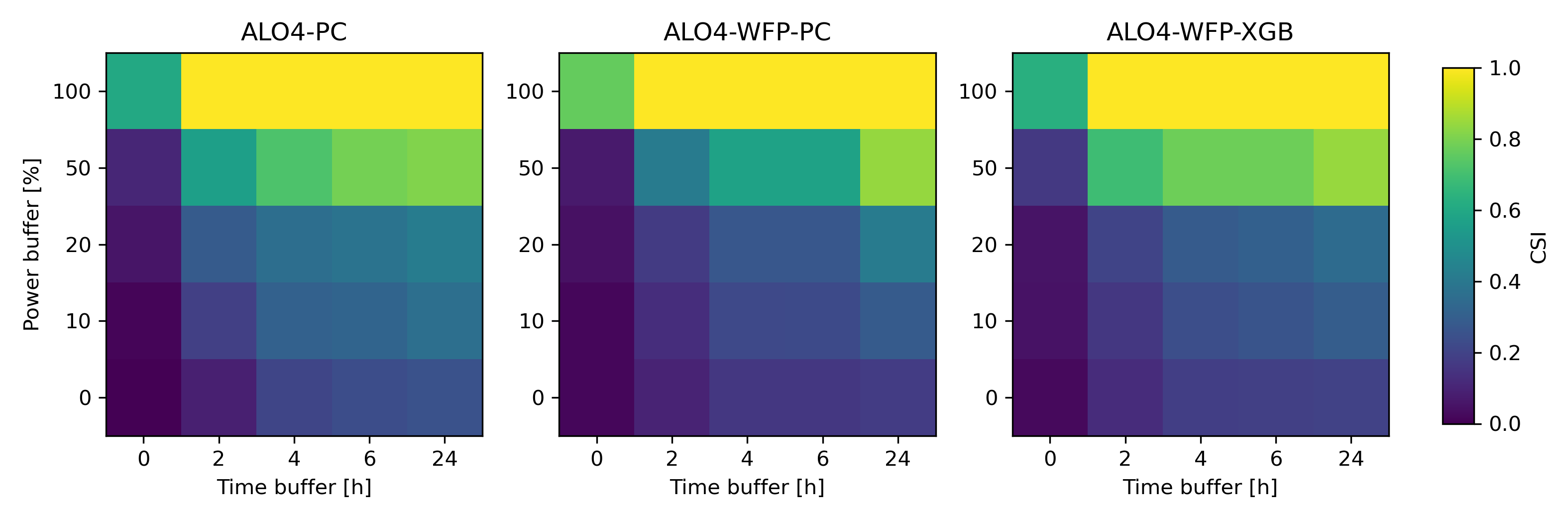} 
    \caption*{Figure \ref{fig:CSI_by_buffer_intraday}A: Day-ahead CSI for hourly 50\% up-ramping events.}
    \label{fig:CSI_by_buffer_dayahead}
\end{figure}

\begin{figure}[ht]
    \centering
    \includegraphics[width=\textwidth]{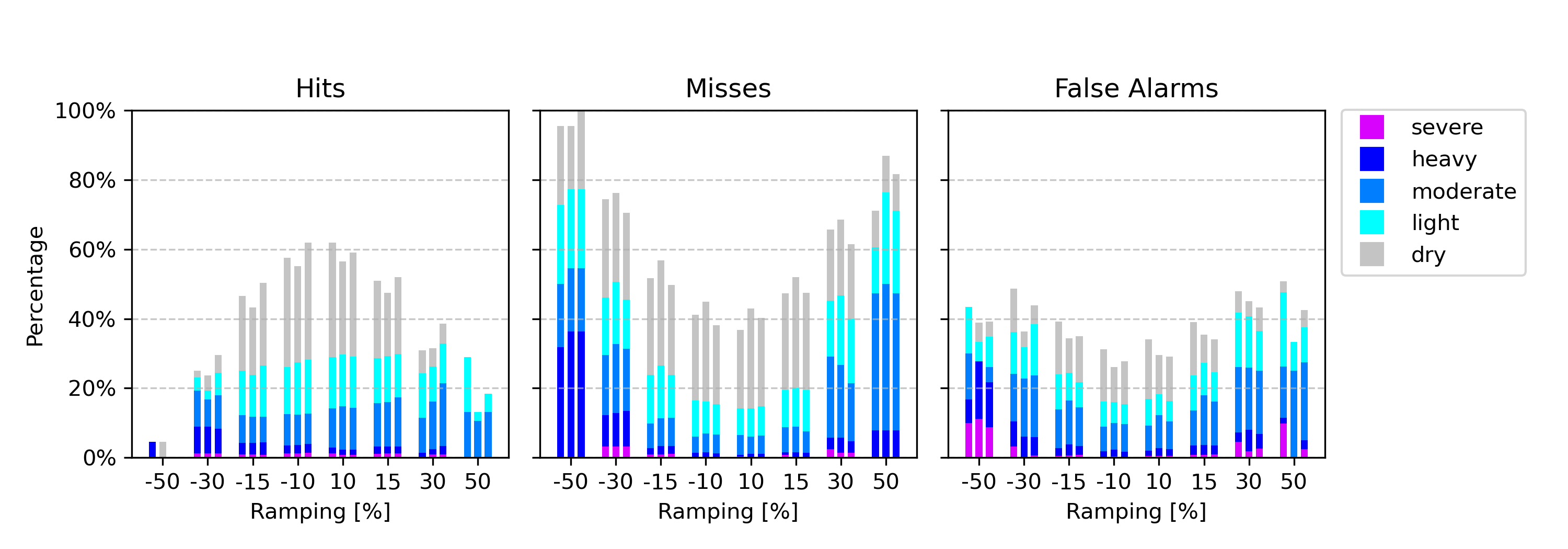}
    \caption*{Figure \ref{fig:dp_PCP_level_intraday}A: Percentage of day-ahead ramping events predictions by hits (left), misses (middle), and false alarms (right).}
    \label{fig:dp_PCP_level_dayahead}
\end{figure}

\clearpage

\end{document}